\documentclass[12pt]{article}
\pdfoutput=1

\usepackage{amsmath}
\usepackage{amssymb}
\usepackage{epsfig}
\usepackage{graphicx}
\usepackage{cite}
\usepackage{multirow}
\usepackage{longtable}
\usepackage{lscape}
\usepackage{bbm}
\usepackage{dcolumn}
\usepackage{rotating}

\allowdisplaybreaks[4] 

\newcommand{\capdef}{}
\newcommand{\mycaption}[2][\capdef]{\renewcommand{\capdef}{#2}%
        \caption[#1]{{\footnotesize #2}}}
\makeatletter
\renewcommand{\fnum@table}{\textbf{\tablename~\thetable}}
\renewcommand{\fnum@figure}{\textbf{\figurename~\thefigure}}
\makeatother

\newcounter{myenumi}

\renewcommand{\themyenumi}{\roman{myenumi}}
{\end{list}}

\newlength{\myem}
\newcounter{mysubequation}[equation]

\makeatletter
\renewcommand{\section}{\@startsection{section}{1}{0em}{-\baselineskip}%
{\baselineskip}{\normalfont\large\bfseries}}
\renewcommand{\subsection}%
{\@startsection{subsection}{2}{0em}{-0.7\baselineskip}%
{0.7\baselineskip}{\normalfont\bfseries}}
\makeatother

\newcommand{\bi}{\begin{itemize}}
\newcommand{\ei}{\end{itemize}}

\newcommand{\be}{\begin{equation}}
\newcommand{\ee}{\end{equation}}
\newcommand{\bea}{\begin{eqnarray}}
\newcommand{\eea}{\end{eqnarray}}

\newcommand{\deltacp}{\delta_{\mathrm{CP}}}

\newcommand{\ie}{{\it i.e.}}

\newcommand{\eg}{{\it e.g.}}

\newcommand{\cf}{{\it cf.}}

\newcommand{\etc}{{\it etc.}}
\newcommand{\eq}{Eq.}

\newcommand{\fig}{Fig.}

\newcommand{\Ref}{Ref.}
\newcommand{\Refs}{Refs.}
\newcommand{\Sec}{Sec.}

\newcommand{\Tab}{Table}

\newcommand{\equ}[1]{\eq~(\ref{equ:#1})}
\newcommand{\figu}[1]{\fig~\ref{fig:#1}}

\begin{document}

\begin{titlepage}

\renewcommand{\thefootnote}{\alph{footnote}}

\vspace*{-3.cm}
\begin{flushright}
IFT-UAM/CSIC-10-43 
\end{flushright}


\renewcommand{\thefootnote}{\fnsymbol{footnote}}
\setcounter{footnote}{-1}

{\begin{center}
{\large\bf
Energy dependent neutrino flavor ratios from cosmic accelerators on the Hillas plot 
} 

\end{center}}

\renewcommand{\thefootnote}{\alph{footnote}}

\vspace*{.8cm}
\vspace*{.3cm}
{\begin{center} {\large{\sc 
                S.~H{\"u}mmer\footnote[1]{\makebox[1.cm]{Email:}
                svenja.huemmer@physik.uni-wuerzburg.de},
		M.~Maltoni\footnote[2]{\makebox[1.cm]{Email:}
                michele.maltoni@uam.es},
                W.~Winter\footnote[3]{\makebox[1.cm]{Email:}
                winter@physik.uni-wuerzburg.de}, and 
                C.~Yaguna\footnote[4]{\makebox[1.cm]{Email:}
                carlos.yaguna@uam.es}
                }}
\end{center}}
\vspace*{0cm}
{\it
\begin{center}

\footnotemark[1]${}^,$\footnotemark[3]
       Institut f{\"u}r Theoretische Physik und Astrophysik, \\ Universit{\"a}t W{\"u}rzburg, 
       97074 W{\"u}rzburg, Germany

\footnotemark[2]${}^,$\footnotemark[4]
      Universidad Autonoma de Madrid and Instituto de F{\'i}sica Te{\'o}rica UAM/CSIC, Cantoblanco, E-28049 Madrid, Spain

\end{center}}

\vspace*{1.5cm}

\begin{center}
{\Large \today}
\end{center}

{\Large \bf
\begin{center} Abstract \end{center}  }

We discuss neutrino fluxes and energy dependent flavor ratios of cosmic accelerators as a function of the size of the acceleration region and the magnetic field, which are the parameters of the Hillas plot.
We assume that 
photohadronic interactions between Fermi accelerated protons and synchrotron photons from co-accelerated electrons (or positrons) lead to charged pion production. We include synchrotron cooling of the charged pions and successively produced muons, which then further decay into neutrinos, as well as the helicity dependence of the muon decays. Our photohadronic interaction model includes direct production, higher resonances, and high energy processes in order to model the neutrino flavor and neutrino-antineutrino ratios with sufficient accuracy. Since we assume that the sources are  optically thin to neutron interactions, we include the neutrino fluxes from neutron decays.  We classify the Hillas plot into regions with different characteristic flavor ratios over 20$\times$24 orders of magnitude in $R$ and $B$. In some examples with sizable magnetic fields, we recover neutron beam, pion beam, and muon damped beam as a function of energy.  However, we also find anomalous or new sources, such as muon beam sources with a flavor ratio $\nu_e$:$\nu_\mu$:$\nu_\tau$ of 1:1:0 in energy regions where the synchrotron-damped muons pile up.  We also discuss the use of the Glashow resonance to identify $p\gamma$ optically thin sources with a strong imbalance between $\pi^+$ and $\pi^-$ production. We find that these can, in principle, be identified in most cases in spite of the $\pi^-$ contamination from high energy photohadronic processes and the mixing parameter uncertainties. 

\vspace*{.5cm}

\end{titlepage}

\newpage

\renewcommand{\thefootnote}{\arabic{footnote}}
\setcounter{footnote}{0}

\section{Introduction}

Neutrino telescopes~\cite{Aslanides:1999vq,Ahrens:2002dv,Tzamarias:2003wd, Piattelli:2005hz}, such as IceCube or ANTARES, are designed to detect neutrinos from astrophysical sources. There are numerous candidate sources, the most prominent extragalactic ones being Active Galactic Nuclei (AGNs)~\cite{Stecker:1991vm,Mannheim:1993,Mucke:2000rn,Aharonian:2002} and Gammay Ray Bursts (GRBs)~\cite{Waxman:1997ti}, see \Ref~\cite{Becker:2007sv} for a review and \Ref~\cite{Rachen:1998fd} for the general theory of the astrophysical neutrino sources. For instance, in GRBs, photohadronic interactions are expected to lead to a significant flux of neutrinos in the fireball scenario. On the other hand, in AGN models, the neutral pions produced in these interactions may describe the second hump in the observed photon spectrum, depending on the dominance of synchrotron or inverse Compton cooling of the electrons.
The protons in these models are typically assumed to be accelerated in the relativistic outflow together with electron and positrons by Fermi shock acceleration. The target photon field is often assumed to be the synchrotron photon field of the co-accelerated electrons and positrons. Furthermore, thermal photons from broad line regions or the accretion disc may serve as target photons, and there may be proton-proton interactions contributing to the neutrino flux. In this work, we focus on sources with a target photon field from synchrotron photons.  
Note that apart from AGNs and GRBs, there a many other candidates for neutrino sources. However, so far, no extraterrestrial high energy neutrino flux has been detected yet. This is, for sources optically thin to neutrons, consistent with generic bounds~\cite{Waxman:1998yy,Mannheim:1998wp} which are just being touched by IceCube. Apart from that, there is also the possibility to detect point sources for which the optical counterpart is absorbed, so-called ``hidden sources''~\cite{Razzaque:2004yv,Ando:2005xi,Razzaque:2005bh,Razzaque:2009kq}. We do not consider particular source properties, but we discuss generic point sources optically thin to neutrons.

Apart from measuring muon tracks, neutrino telescopes may have some sensitivity to detect flavor~\cite{Learned:1994wg,Beacom:2003nh}. The neutrinos are normally assumed
to originate from pion decays, with a flavor ratio $\nu_e$:$\nu_\mu$:$\nu_\tau$  at the source of 1:2:0 (neutrinos and antineutrinos added) arising from the
decays of both primary pions and secondary muons. However, it was pointed out in \Refs~\cite{Rachen:1998fd,Kashti:2005qa} that energy losses in strong magnetic fields, which
dominantly affect the muons, change the flavor ratio at the source to 0:1:0. At low energies, neutron decays may dominate the flux, which leads to a flavor ratio 1:0:0. Therefore, one
can expect a smooth transition from one type of source to the other as
a function of the neutrino energy~\cite{Kachelriess:2006fi,Lipari:2007su,Kachelriess:2007tr}, depending
on the cooling processes of the intermediate muons, pions, and kaons. Recently,
the use of flavor information has been especially proposed to extract some information on the particle physics properties of the neutrinos and the properties of the source; see \Ref~\cite{Pakvasa:2008nx} for a review.
For instance, if the neutrino telescope has
some flavor identification capability, such that it can distinguish between muon tracks and  (electromagnetic or hadronic) showers, this property can be used to
extract information on the decay~\cite{Beacom:2002vi, Lipari:2007su,Majumdar:2007mp,Maltoni:2008jr,Bhattacharya:2009tx,Bhattacharya:2010xj}
and oscillation~\cite{Farzan:2002ct, Beacom:2003zg,Serpico:2005sz, Serpico:2005bs,Bhattacharjee:2005nh, Winter:2006ce, Majumdar:2006px,Meloni:2006gv, Blum:2007ie, Rodejohann:2006qq, Xing:2006xd, Pakvasa:2007dc, Hwang:2007na, Choubey:2008di,Esmaili:2009dz} parameters. Of course, the flavor ratios may be also used for source identification, see, \eg, \cite{Xing:2006uk,Choubey:2009jq}. Except from flavor identification,
the differentiation between neutrinos and antineutrinos could be useful for the
discrimination between $p \gamma$ and $pp$ induced neutrino fluxes, or for the
test of neutrino properties (see, \eg, \cite{Maltoni:2008jr}). A useful
observable may be the Glashow resonance process $\bar{\nu}_e + e^- \to W^- \to
\text{anything}$ at around $6.3 \, \text{PeV}$~\cite{Learned:1994wg,Anchordoqui:2004eb, Bhattacharjee:2005nh}
to distinguish between neutrinos and antineutrinos in the detector. Note that for
photohadronic interactions, the $\pi^+$ to $\pi^-$ ratio determines the ratio between
electron neutrinos and antineutrinos at the source. 

In the existing literature, the flavor ratio is often taken to be a constant number independent of energy, with the above mentioned standard values taken as dogma. We demonstrate that even for ``standard'' sources, such as AGNs, flavor ratios in more realistic models could be anomalous in the sense that they do not necessarily fit the above picture. However, if the astrophysical parameters of the source, such as the extension of the acceleration region, are measured otherwise, such as by the variability timescale of the photon field flux, the energy dependence can be predicted in simple models. The idea of our model it to define the simplest possible toy model including a synchrotron target photon field and magnetic field effects on the secondaries. Naturally, it cannot describe all types of sources accurately. On the other hand, it does not rely on any astrophysical empirical relationships. 
Our main parameters are a universal injection index $\alpha$ for protons and electrons/positrons, the size of the acceleration region $R$ and the magnetic field $B$. Our model is self-consistent in the sense that the target photon field is not put in by hand, but assumed to come from the synchrotron emission from electrons and positrons co-accelerated with the protons. Therefore, it relies on relatively few parameters (plus some assumptions). Pions are produced in photohadronic interactions between the protons and synchrotron photons, which leads to pion (and kaon) fluxes, and finally to neutrino fluxes. We include flavor and magnetic field effects, which means that we keep track of all flavors separately, include the magnetic field effects on pions, muons, and kaons, as well as the helicity dependence of the muon decays. The source is assumed to be optically thin to neutrons produced by photohadronic interactions, which means that we have to include the electron antineutrinos from neutron decays.
Note that because we are only interested in the neutrino spectra, which are measured with too little statistics to exhibit time dependent features, we choose a time independent (steady state) approach. 
A similar self-consistent model can, for the specific case of microquasars, be found in \Ref~\cite{Reynoso:2008gs}. For other refined neutrino flux calculations, see, \eg, \Refs~\cite{Lipari:2007su,Mucke:2000rn,Muecke:2002bi,Muecke:1999ICRC}

The parameters $R$ and $B$ can be directly related to the Hillas plot. In order to confine a particle in a magnetic field at the source, the Lamor radius has to be smaller than the extension of the acceleration region $R$. This can be translated into the Hillas condition for the maximal energy~\cite{Hillas:1985is}
\begin{equation}
E_{\mathrm{max}} \, [\mathrm{GeV}] \simeq 0.03 \cdot \eta \cdot Z \cdot  R \, [\mathrm{km}] \cdot B \, [\mathrm{G}] \, .
\label{equ:hillas}
\end{equation}
Here $Z$ is the charge (number of unit charges) of the accelerated particle, $B$ is the magnetic field in Gauss, and $\eta$ can be interpreted as an efficiency factor or linked to the characteristic velocity of the scattering centers. In this study, we consider the parameters $R$ and $B$ varied on the Hillas plot over 20$\times$24 orders of magnitude. 

Since we study neutrino fluxes, flavor ratios, and neutrino-antineutrino ratios, we need a sufficiently fast model for the photomeson production which reproduces these features for power law-like spectra quite accurately. The often used $\Delta$-resonance approximation
\begin{equation}
p + \gamma \rightarrow \Delta^+ \rightarrow \left\{
\begin{array}{ll}
n + \pi^+ & 1/3 \, \, \text{of all cases} \\
p + \pi^0 & 2/3 \, \, \text{of all cases} \\
\end{array}
\right. .
\label{equ:ds}
\end{equation}
is not sufficient for this purpose, since, for example, the $\pi^-$ contamination is not included, and high energy processes affecting the spectral shapes are not treated at all.  The best option may be the SOPHIA software~\cite{Mucke:1999yb}, a Monte Carlo simulation including not only the full final state kinematics, but also higher resonances, multi-pion production, and direct ($t$-channel) production of pions. This option is, however, too slow for our large scale parameter space scan, especially if smooth flavor ratio predictions are needed, and does not contain the helicity dependence of the muon decays, which is important for flavor ratio computations~\cite{Lipari:2007su}. Therefore, we use \Ref~\cite{Hummer:2010vx} (model Sim-B) including the helicity dependent muon decays from \Ref~\cite{Lipari:2007su}. This approach is largely based on the physics of SOPHIA. However, the kinematics are simplified, which can be seen mainly for sharp features in the proton and photon spectra at high energies, whereas for power law spectra, the treatment is very accurate. Compared to the approach in \Ref~\cite{Kelner:2008ke}, which approximates the SOPHIA treatment analytically, the intermediate particles (pions, muons, kaons) are not integrated out, and magnetic field effects relevant for the flavor ratios can be included. For any given set of model parameters, our C-software NeuCosmA (``Neutrinos from Cosmic Accelerators'') can compute all secondary and neutrino flux spectra for one set of parameters in only a few seconds. This, of course, allows for applications which have not been possible in the past, such as the Hillas space scans in this study. In principle, our model can also be used as a (almost) online fit model for fluxes, because it interpolates among wide parameter space regions.

This study is organized as follows: In \Sec~\ref{sec:model}, we describe our cosmic accelerator model, show a typical example, and discuss its limitations. The reader familiar with similar models may at least want to read \Sec~\ref{sec:summarymodel}. In \Sec~\ref{sec:source}, we discuss the source classification in the shock rest frame, \ie, without Lorentz boost, redshift, and flavor mixing.  In \Sec~\ref{sec:detector}, we then discuss the physics at the detector. We finally summarize in \Sec~\ref{sec:summary}.

\section{The source model}
\label{sec:model}

Here we discuss our model for the neutrino source. A (qualitative) model summary can be found in \Sec~\ref{sec:summarymodel}, whereas a more quantitative component description can be found in \Sec~\ref{sec:description}. Then we discuss typical qualitative and quantitative results in \Sec~\ref{sec:typresults}. In \Sec~\ref{sec:limitations}, we comment on the limitations of our model.

\subsection{Short model summary}
\label{sec:summarymodel}

We assume that protons and electrons (or positrons) are co-accelerated by Fermi processes in the source such that their injection spectra are power laws with the universal injection index $\alpha$. The synchrotron radiation of the electrons in the magnetic field $B$ of the source, assuming an isotropic pitch angle distribution, is the considered target photon spectrum for the protons. Photohadronic interactions lead to charged pion, kaon, and neutron production. These particles decay further through weak decays into muons and neutrinos. 
Our photohadronic interaction model is based on \Ref~\cite{Hummer:2010vx} (model Sim-B) including direct production, different resonances, and high energy processes. The helicity dependent muon decays are taken into account as in \Ref~\cite{Lipari:2007su}.
The resulting neutrino spectra have to be Lorentz boosted (and redshifted) to obtain the spectra at the Earth. We assume that the source is optically thin to neutrons, which means that  neutrons produced by the photohadronic interactions will decay via beta decay and lead to an additional electron antineutrino flux. 
For all charged particles we include the effect of adiabatic and synchrotron energy losses, and we assume that these dominate such that energy losses due to inverse Compton processes, Bethe-Heitler pair production, and photohadronic interactions (the source is optically thin to neutrons) are sub-leading. Although not all spectral features of some sources may be described by this method, it has the advantage that the electron-positron (or proton) ratio is not needed as parameter, and that only the product of electron and proton injection spectrum enters the total normalization.

\begin{figure}[t]
 \includegraphics[width=1\textwidth,viewport=20 192 696 548]{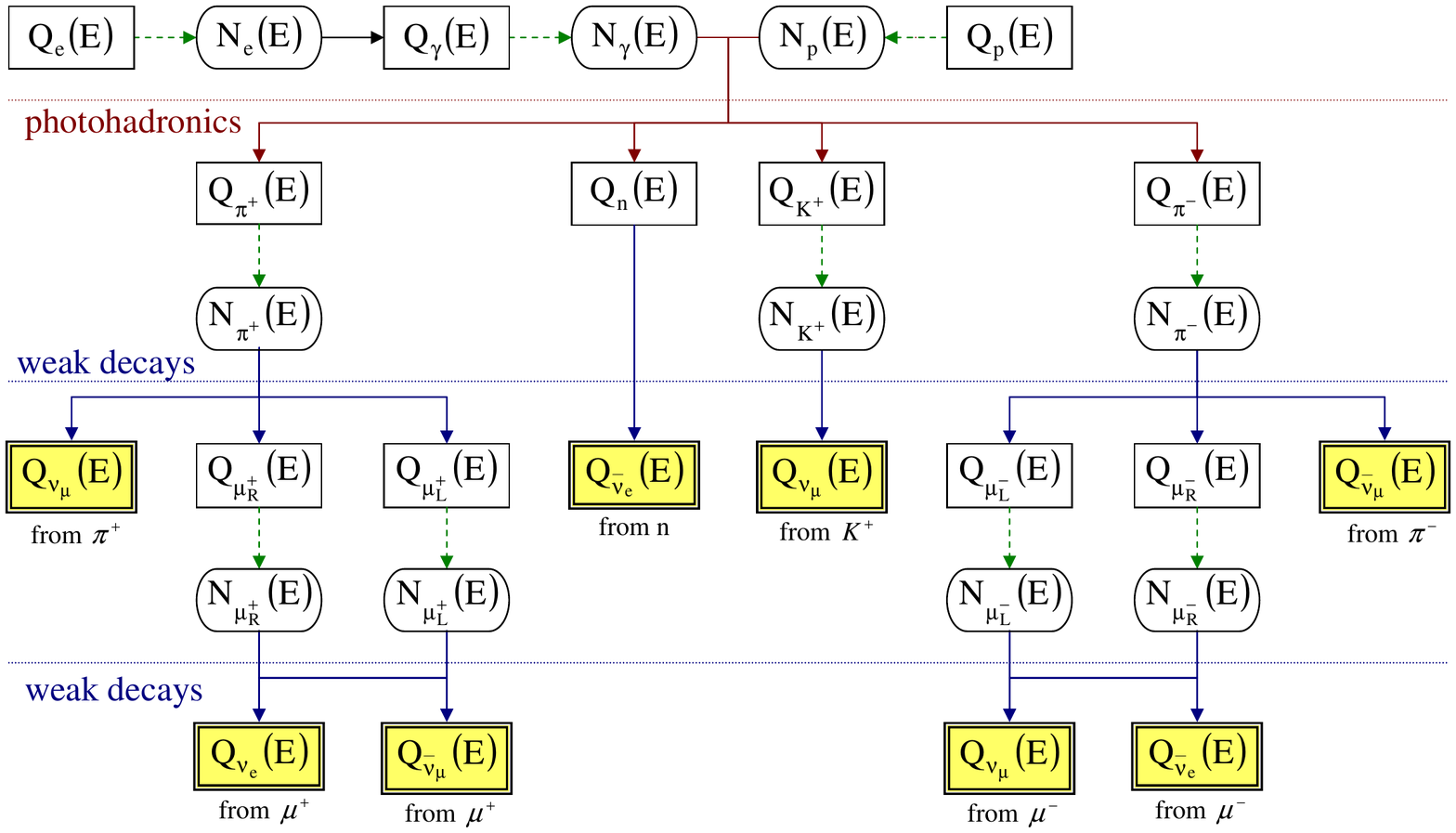}
 \mycaption{\label{fig:flowchart}Flowchart describing the model. The functions $Q(E)$ denote (injection) spectra per time frame $[\mathrm{\left( GeV\,s\,cm^3\right)^{-1}}]$ and $N(E)$ steady spectra $[\mathrm{\left( GeV\,cm^3\right)^{-1}}]$ derived from the balance between injection and losses or escape. Dashed arrows stand for solving the steady state differential equation \equ{steadstate}.}
\end{figure}

To summarize, we show in  \figu{flowchart} a flowchart describing the  computation of the neutrino spectra in the shock rest frame (SRF). Here $Q(E)$ (rectangular boxes) stand for (injection) spectra per time frame and $N(E)$ (boxes with rounded corners) for steady spectra derived from the balance between injection and losses or escape. Dashed arrows stand for solving the steady state differential equation balancing energy losses and escape with injection. Weak decays and photohadronic interactions are denoted by the horizontal lines. 
The resulting neutrino spectra are marked by a double border and a colored background. Below the boxes we indicate in which decay the neutrinos are produced.

\begin{table}[t]
\begin{center}
\begin{tabular}{llll}
\hline
Parameter & Units & Description & Typical values used \\
\hline
$R$ & km (kilometers)&  Size of acceleration region & $10^1 \, \mathrm{km} \hdots 10^{21} \, \mathrm{km}$ \\
$B$ & G (Gauss) & Magnetic field strength & $10^{-9} \, \mathrm{G} \hdots 10^{15} \, \mathrm{G}$ \\
$\alpha$ & 1 & Universal injection index & $1.5 \hdots 4$ \\
$E_{e,\mathrm{min}}$ & GeV & Minimal electron energy & $m_e$ \\
$E_{p,\mathrm{min}}$ & GeV & Minimal proton energy & $m_p$ \\
$\eta$ & 1 & Acceleration efficiency & 0.1 \\
$\Gamma$ & 1 & Lorentz boost factor & $1 \hdots 1000$ \\
$N$ & a.u. & Normalization/luminosity & arbitrary units \\
\hline
\end{tabular}
\end{center}
\mycaption{\label{tab:params} Parameters of our model, units, description, and typical values used in this study. }
\end{table}

The parameters of the source are all defined in the SRF. We list them in \Tab~\ref{tab:params}, together with units and typically used parameter values. The above mentioned magnetic field  $B$ is important for the synchrotron radiation and the related energy losses. The size of the acceleration region $R$ plays an important role for the description of adiabatic losses and for the escape of particles leaving the source. The maximal energy of injected electrons and protons is limited where the energy loss rate exceeds the acceleration rate, which includes the parameter $\eta$ (acceleration efficiency) from \equ{hillas}. The universal injection index $\alpha$ describes the index of the injection spectrum of protons and electrons. To fully describe these injection spectra the minimal energies of the proton ($E_{p,\mathrm{min}}$) and electron ($E_{e,\mathrm{min}}$) spectra and their normalizations have to be given. Our model is constructed such that only the product of these normalizations enters in the total normalization of the neutrino  spectrum. Since we mainly discuss the spectra shapes and flavor ratios of the neutrino fluxes, we typically do not consider the normalization as separate parameter. The last parameter is a possible  Lorentz factor $\Gamma$ of a relativistic outflow which allows us to boost the neutrino spectra into the observer's frame at the Earth. We only consider this parameter in \Sec~\ref{sec:detector}. The effects due to the cosmological redshift of the source are not explicitely discussed since they can be absorbed in a more generally defined Doppler factor (\cf, \Ref~\cite{Rachen:1998fd}).

\subsection{Components of the model}
\label{sec:description}

Here we quantitatively describe the key ingredients of our model. Unless noted otherwise, all quantities are given in the SRF.

\subsubsection*{Synchrotron target photon field}

We calculate the synchrotron spectrum of the electrons in the Melrose-approximation~\cite{Melrose:1980gb} averaged over the pitch angle. The power radiated per photon energy $\varepsilon$ by one particle with energy $E$, mass $m$ and charge $q$ in a magnetic field $B$ is then given by:
\begin{equation}
\label{equ:singlesyn}
 P_s(\varepsilon,E)=1.8\cdot\frac{\sqrt{3}\,q^3\,B}{16\,\varepsilon_0\,m\,c\,h}\cdot\left(\frac{\varepsilon}{\varepsilon_c}\right)^{1/3}e^{-\varepsilon/\varepsilon_c}\quad\text{with}\quad \varepsilon_c=\frac{3\,q\,B}{16\,m}\left(\frac{E}{m\,c^2}\right)^2.
\end{equation}
Considering a spectrum of particles radiating in a magnetic field we have to convolute  with the spectrum of radiating particles as
\begin{equation}
 P(\varepsilon)=\int_0^\infty dE \, N(E)\, P_s(\varepsilon,E) \, .
\end{equation}
The number of produced photons per time can be computed with:
\begin{equation}
 Q_\gamma(\varepsilon)=\frac{P(\varepsilon)}{\varepsilon}.
\end{equation}
The steady photon spectrum, which is needed for the computation of photohadronic interactions, can be estimated by multiplying $Q_\gamma$ in units
 $\mathrm{\left( GeV\,s\,cm^3\right)^{-1}}$ with the escape time $t_{\mathrm{esc}} \simeq R/c$ of the photons. The photon spectrum $N_\gamma$ is then given by
\begin{equation}
\label{equ:photonnorm}
 N_\gamma(\varepsilon)=Q_\gamma(\varepsilon) \, \frac{R}{c}
\end{equation}
in units of $\mathrm{\left( GeV\,cm^3\right)^{-1}}$. In the following, $Q(E)$ will always denote spectra per time frame in units of $\mathrm{\left( GeV\,s\,cm^3\right)^{-1}}$, whereas  $N(E)$ will refer to steady spectra in units of $\mathrm{\left( GeV\,cm^3\right)^{-1}}$. 

\subsubsection*{Photohadronic interactions}

The accelerated protons interact with the synchrotron photons radiated by the electrons. In these interactions, pions, kaons, and neutrons are produced. We consider the following processes:
\begin{align}
 p + \gamma & \rightarrow \pi + p' \, ,\\
 p + \gamma & \rightarrow K^+ + \Lambda/\Sigma \, .
\end{align}
Here $p'$ is either a proton or neutron, $\pi$ are one or more neutral, positive, or negative charged pions, and $\Lambda$ and $\Sigma$ are resonances.
Note that the effect of kaon decays is usually small. However, kaon decays may have interesting consequences for the neutrino flavor ratios at very high energies, in particular, if strong magnetic fields are present (the kaons are less sensitive to synchrotron losses because of their larger rest mass)~\cite{Kachelriess:2006fi,Kachelriess:2007tr}.
Therefore, we consider the leading mode: $K^+$ production (for protons in the initial state) with the decay channel leading to highest energy neutrinos.\footnote{The contributions from $K^-$ and $K^0$ are about a factor of two suppressed~\cite{Lipari:2007su}, and $K^+$ has a leading two body decay mode into neutrinos.} Note that at even higher energies, other processes, such as charmed meson production, may contribute as well.

We follow the description of the photohadronic interactions in \Ref~\cite{Hummer:2010vx} (Sim-B), based on the physics of SOPHIA~\cite{Mucke:1999yb}. The treatment, taking into account resonant (including higher resonances), direct, and high energetic multi-pion production, allows for an efficient computation which is rather precise for power law-like spectra. It predicts well neutrino to antineutrino ratios and flavor ratios. The interaction between an arbitrary proton spectrum $N_p(E_p)$ and an arbitrary target photon field $N_\gamma(\varepsilon)$ is described by
\begin{equation}
Q_b(E_b) = \int\limits_{E_b}^{\infty} \frac{dE_p}{E_p} \, N_p(E_p) \, \int\limits_{\frac{\epsilon_{\mathrm{th}} m_p}{2 E_p}}^{\infty} d\varepsilon \, N_\gamma(\varepsilon) \,  R_b( x,y )  
\label{equ:prodmaster}
\end{equation}
with $x=E_b/E_p$ the fraction of energy going into the secondary, $y \equiv (E_p\varepsilon)/m_p$ (directly related to the center of mass energy), the response function $R_b( x,y )$, and $\epsilon_{\mathrm{th}}$ the threshold for the photohadronic interactions. In our calculations, we use model Sim-B, which is characterized by the factorized response function
\begin{equation}
R_b^{\mathrm{IT}}(x,y)= \delta(x - \chi^{\mathrm{IT}}) \,  M_b^{\mathrm{IT}} \, f^{\mathrm{IT}}(y)  \quad \text{with} \quad
  f^{\mathrm{IT}}(y) \equiv \frac{1}{2y^2} \int\limits_{\epsilon_{\mathrm{th}}}^{2y} d \epsilon_r \, \epsilon_r \, \sigma^{\mathrm{IT}}(\epsilon_r)  \, 
\label{equ:split}
\end{equation}
summed, for instance, over 23 different interaction types (IT) per pion species. For details, see \Ref~\cite{Hummer:2010vx}. With the factorization of the response function only a single integration is necessary to determine the spectra of the produced secondaries. This fact allows for a fast computation.
From \equ{prodmaster}, it is also easy to see that in $Q_b(E)$ only the product of the normalizations of the proton and photon (radiated from electrons) fluxes enters.

\subsubsection*{Weak decays}

We let the produced particle species, namely pions, kaons, and neutrons, decay weakly into neutrinos as
\begin{eqnarray}
\pi^+ & \rightarrow & \mu^+ + \nu_\mu \, ,\nonumber \\
& & \mu^+ \rightarrow e^+ + \nu_e + \bar{\nu}_\mu \, , \label{equ:piplusdec} \\
\pi^- & \rightarrow & \mu^- + \bar\nu_\mu \, , \nonumber \\
& & \mu^- \rightarrow e^- + \bar\nu_e + \nu_\mu \, , \label{equ:piminusdec} \\
 K^+ & \rightarrow & \mu^+ + \nu_\mu  \label{equ:kplusdec} \, ,\\
 n & \rightarrow & p + e^- + \bar{\nu}_e  \label{equ:ndec}  \, .
\end{eqnarray}
Note again that for kaons, the branching ratio for this channel is about 64\%.
The second-most-important decay mode is $K^\pm \rightarrow \pi^\pm +  \pi^0$ (20.7\%).
The other decay modes account for 16\%, no more than about 5\% each.
Since interesting effects can only be expected in the energy range with the most energetic neutrinos, we only use the direct decays from the leading mode.

Since the decays of muons are helicity dependent, we distinguish between left and right handed muons (see \Ref~\cite{Lipari:2007su}). In general, in case of ultra-relativistic parents of type $a$, the distribution of the daughter particle of type $b$ takes a scaling form in order to obtain for the energy spectra
\begin{equation}
 Q_b(E_b)=\sum_a\int_{E_b}^\infty\,dE_a\,N_a(E_a)\,t_\mathrm{dec}^{-1}\,\frac{1}{E_a}\,F_{a\rightarrow b}\left(\frac{E_b}{E_a}\right) 
\end{equation}
summed over all parent species.
 The functions $F_{a\rightarrow b}$ for pion, kaon and helicity dependent muon decays can be read off from \Ref~\cite{Lipari:2007su} (Sec. IV). For decays of neutrons produced by photohadronic interactions we use (\cf, \Ref~\cite{Lipari:2007su}):
\begin{equation}
 F_{n\rightarrow \bar{\nu}_e}=\delta\left(\frac{E_\nu}{E_n}-\chi_n \right).
\end{equation}
with $\chi_n=5.1\times10^{-4}$. Since our source is assumed to be optically thin to neutrons and the neutron lifetime is relatively short, we assume that any produced neutron will decay between production and arrival at the Earth, either inside or outside the source. Thus, the neutrino spectrum can be directly related to the neutron injection spectrum as
\begin{equation}
 Q_{\bar{\nu}_e}(E_\nu)=\frac{1}{\chi_n} Q_n\left(\frac{E_\nu}{\chi_n}\right) \, .
\label{equ:ndecay}
\end{equation}

\subsubsection*{Energy losses and escape}

For the charged species we consider possible energy losses, such as synchrotron radiation, before they decay. The kinetic equation for the particle spectrum, assuming continuous energy losses, is given by (see, \eg , \Ref~\cite{Atoyan:2002gu}):
\begin{equation}
\frac{\partial N(E)}{\partial t}=\frac{\partial}{\partial E}\left(-b(E) \, N(E)\right)-\frac{N(E)}{t_\mathrm{esc}}+Q(E)
\end{equation}
with $t_\mathrm{esc}(E)$ the characteristic escape time, $b(E)=-E\,t_\mathrm{loss}^{-1}$ with $t_\mathrm{loss}^{-1}(E)=-1/E \, dE/dt$ the rate characterizing energy losses, $Q(E)$ the particle injection rate $[\mathrm{\left( GeV\,s\,cm^3\right)^{-1}}]$ 
and $N(E)$ the steady particle spectrum $[\mathrm{\left( GeV\,cm^3\right)^{-1}}]$. For different energy loss or decay mechanisms, the inverse timescales are summed over. Because the statistics in neutrino observations is not expected to be  high enough for resolving time dependent features, we assume that energy losses and escape are in balance with the particle injection and that $\partial N(E)/\partial t=0$, which leads to the so-called steady state spectrum $N(E)$ described by the differential equation
\begin{equation}
\label{equ:steadstate}
Q(E)=\frac{\partial}{\partial E}\left(b(E) \, N(E)\right)+\frac{N(E)}{t_\mathrm{esc}} \, .
\end{equation}
This steady state is another key ingredient for the efficient computation of neutrino fluxes, since time dependent calculations, which are the state-of-the-art for photon flux predictions, are extremely time intensive.
Let us now discuss the special case of no energy losses, such as for photons, \ie,  $b(E)=0$. Then we can easily solve the differential equation in \equ{steadstate} in order to obtain 
\begin{equation}
 \label{eqn:steadystateesc}
 N(E)=Q(E)\,t_\mathrm{esc} \, ,
\end{equation}
which is just the same equation as \equ{photonnorm}.

\begin{table}[t]
\begin{center}
   \begin{tabular}[t]{l|cc|cc}
  \hline
   & \multicolumn{2}{c|}{Energy loss}  & \multicolumn{2}{c}{Escape} \\
   \hline 
   Particle species & Synchrotron & Adiabatic & Escape (region) & Decay \\ 
   \hline
   Electron/proton & x & x &  &  \\
   Target photons &  &  & x & \\
   $\pi^-$/$\pi^+$/$K^+$/$\mu^-$/$\mu^+$ & x & x &  &  x \\
   \hline
  \end{tabular}
 \mycaption{\label{tab:timescales} Considered energy loss and escape timescales in the steady state for the considered particle species.}
\end{center}
\end{table}

We summarize the considered energy loss and decay timescales for the different particle species in \Tab~\ref{tab:timescales}. In the model, we consider the timescales for synchrotron (charged particles) and adiabatic (particles coupling to the plasma) energy losses and for escape from the region (neutral particles) and decays (unstable particles). The timescales can be approximated by the following formulas:
\begin{equation}
 t^{-1}_\mathrm{synchr}=\frac{q^4\,B^2\,E}{9\,\pi\,\epsilon_0\,m^4\,c^5}\ , \quad t^{-1}_\mathrm{ad}\approx\frac{\beta_\mathrm{ex} c}{R},\quad t^{-1}_\mathrm{es}=\frac{c}{R}\ , \quad t^{-1}_\mathrm{decay}=\frac{m c^2}{E\,\tau_0}
\end{equation}
with $q$, $E$, $m$ and $\tau_0$ the charge, energy, mass, and rest frame lifetime of the particle, $B$ the magnetic field, and $R$ the extension  of the acceleration region. The expansion beta factor in the comoving frame $\beta_\mathrm{ex}$ is approximated by one, which is a reasonable assumption for AGNs and GRBs (see \Ref~\cite{Mannheim:1998wp}). 

\subsubsection*{Injection spectrum and maximal energies}

The explicit form of our injection spectra for protons and electrons is:
\begin{align}
 Q_p(E_p)&=\begin{cases}
           N_p\ E^{-\alpha} & E_{p,\mathrm{min}}\,<\,E_p\,<\,E_{p,\mathrm{max}}\\
	   0 & \mathrm{else} \, . 
          \end{cases}\label{equ:pinjection} \\
 Q_e(E_e)&=\begin{cases}
           N_e\ E^{-\alpha} & E_{e,\mathrm{min}}\,<\,E_e\,<\,E_{e,\mathrm{max}}\\
	   0 & \mathrm{else} \, .
          \end{cases} \label{equ:einjection} 
\end{align}
Apart from the injection index $\alpha$ and the two model parameters $E_{p,\mathrm{min}}$ and $E_{e,\mathrm{min}}$, the maximal energies $E_{p,\mathrm{max}}$ and $E_{e,\mathrm{max}}$ have to be determined.
We can estimate these energies by equalizing the characteristic timescale for acceleration of the particles with their energy loss timescale. For this computation, we use the acceleration timescale approximation~\cite{Hillas:1985is}:\footnote{For a more refined discussion, see, \eg, \Ref~\cite{Medvedev:2003sx}. Our assumption corresponds to the ``efficient acceleration'' regime therein. Although there could be additional constraints for  the maximum energies for particles leaving the source from diffuse acceleration processes, for neutrino production, this discussion may only partially apply (the particles producing neutrinos interact before they leave the source).}
  \begin{equation}
  t^{-1}_\mathrm{acc}=\eta\frac{c^2\,Z\,e\,B}{E}
 \end{equation}
with $\eta$ an acceleration efficiency which depends on the acceleration mechanism, and $Z$ the number of unit charges (see also \equ{hillas}).

\subsection{Typical results}
\label{sec:typresults}

In this section, we first discuss what we can qualitatively learn about the secondary and neutrino spectra, and then we show a typical quantitative result.

\subsubsection*{Qualitative shape of the neutrino spectra}

 We have a spectral index $\alpha$ for the injection spectrum of the protons and electrons (in the following, the spectral index is always chosen positive). For large enough energies, the electron spectrum will be loss-steepened by one power due to synchrotron losses. Therefore, the index of the steady state spectrum of protons is $\alpha_p^s=\alpha$ and of electrons $\alpha_e^s=\alpha+1$. Subsequently, we obtain an index  $\alpha_\gamma=(\alpha_e^s-1)/2+1=\alpha/2+1$ for the photon injection spectrum $Q_\gamma(E_\gamma)$. The spectral index of pions produced by photohadronics can be estimated by $\alpha_\pi=\alpha_p^s-\alpha_\gamma+1$. Hence, we obtain $\alpha_\pi=\alpha/2$. If the secondaries, such as pions and muons, do not loose energy, the daughter particles in weak decays have the same spectral index as the parent particles. Therefore, one obtains for the neutrino spectral index
\begin{equation}
 \alpha_\nu=\frac{\alpha}{2}.
\end{equation}
If there are synchrotron energy losses of secondaries, their (steady state) spectrum is loss-steepened by two powers above the critical energy
\begin{equation}
\label{equ:Ecrit}
E_c=\sqrt{\frac{9\,\pi\,\epsilon_0\,m^5\,c^5}{\tau_0\,q^4\,B^2}} \, 
\end{equation}
which is defined as the energy where energy loss time and decay time are equal. This means that for energies higher than the critical energy the particles loose energy before decaying.  \equ{Ecrit} is valid for all charged secondaries with their corresponding masses, charges and lifetimes. Due to their shorter lifetime and their larger mass, the pions critical energy is 18 times higher than for muons. The spectral index for neutrinos from pion $\alpha_\nu^\pi$ and muon $\alpha_\nu^\mu$ decay respectively, is then given by:\footnote{For simplicity, we omit the fact that the neutrinos receive about one fourth (third) of the pion (muon) energy.}
\begin{equation}
 \alpha_\nu^\pi=\begin{cases}
                 \frac{\alpha}{2} & E_\nu\,<\,E_c^\pi\\
		 \frac{\alpha}{2}+2 & E_\nu\,>\,E_c^\pi
                \end{cases}
\, , \qquad
 \alpha_\nu^\mu=\begin{cases}
                 \frac{\alpha}{2} & E_\nu\,<\,E_c^\mu\\
		 \frac{\alpha}{2}+2 & E_\nu\,>\,E_c^\mu \quad .\\
		 \frac{\alpha}{2}+4 & E_\nu\,>\,E_c^\pi
                \end{cases}
\label{equ:coolana}
\end{equation}

\begin{figure}[t]
 \includegraphics[width=1\textwidth]{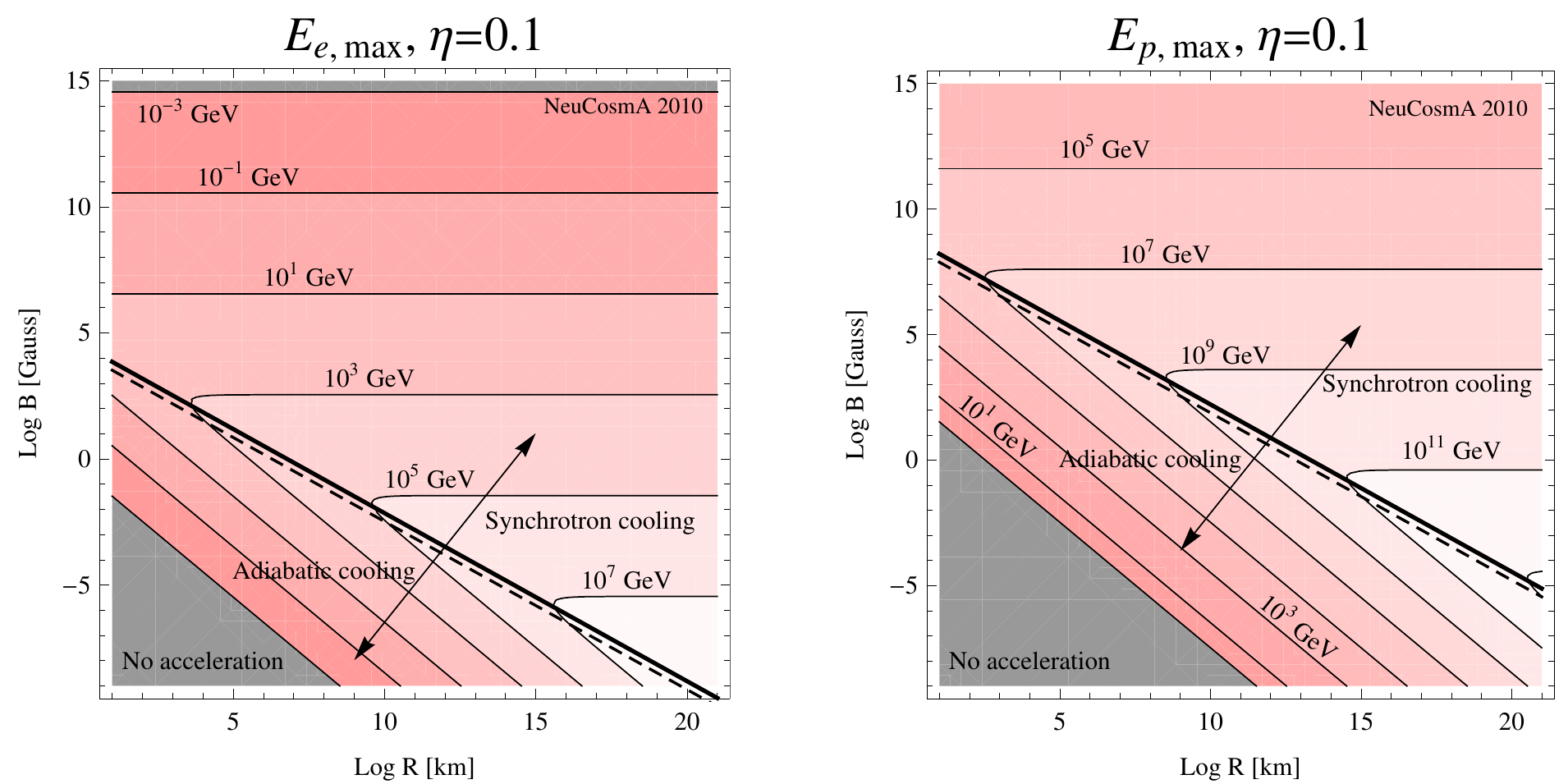}
\mycaption{\label{fig:maxenergy} Maximal energy of electrons (left panel) and protons (right panel) in our model as a function of $R$ and $B$ for acceleration efficiency $\eta=0.1$. The solid (dashed) lines show the border between synchrotron loss and adiabatic loss determined maximal energy for $\eta=0.1$ ($\eta=1$). }
\end{figure}

The discussion of the spectral index for neutrino spectra applies to the energy range in which photohadronic interactions happen. From the maximal energy of the electrons follows the maximal energy of the photons $\varepsilon_\text{max}$. This energy defines the minimal energy of the protons which still undergo photohadronic interactions, denoted by $E^{p\gamma}_\text{min}$, different to the parameter $E_{p,\mathrm{min}}$ describing the minimal energy of the proton spectrum (see \equ{pinjection}). Since the neutrinos obtain approximatively $1/20$ of the proton energy, the index of the neutrino spectrum changes for energies below $E_\nu\lesssim E_\text{min}^{p\gamma}/20$. The maximal neutrino energy can also be estimated by the maximal energy of the protons to $E_\text{max}^\nu\simeq E_{p,\text{max}}/20$. 

In Fig.~\ref{fig:maxenergy} we show the maximal energy of electrons (left panel) and protons (right panel) as a function of the radius $R$ and magnetic field $B$ for the acceleration efficiency $\eta=0.1$. These energies are obtained from the balance between acceleration rate and loss rate. 
In the parameter region above the black solid lines synchrotron losses are dominant for the determination of the maximal energy, whereas below the lines adiabatic losses dominate. The dashed line corresponds to the solid line for $\eta=1$. Changing in addition $\eta$ would change the maximal energy. In the region where synchrotron losses dominate the maximal energy, it is proportional to $\sqrt{\eta}$, and in the adiabatic regime it is proportional to $\eta$. In general, it is clear that protons  can be accelerated to higher energies than electrons due to their higher mass. Note again that it is well known that the maximum particle energies are limited by synchrotron losses in strong magnetic fields, see, \eg, \Refs~\cite{Medvedev:2003sx,Ptitsyna:2008zs}. However, the exact shape of the regions in \figu{maxenergy} depends on the assumptions for the acceleration mechanism. 

\subsubsection*{A quantitative example}

\begin{figure}[t]
     \includegraphics[width=1\textwidth]{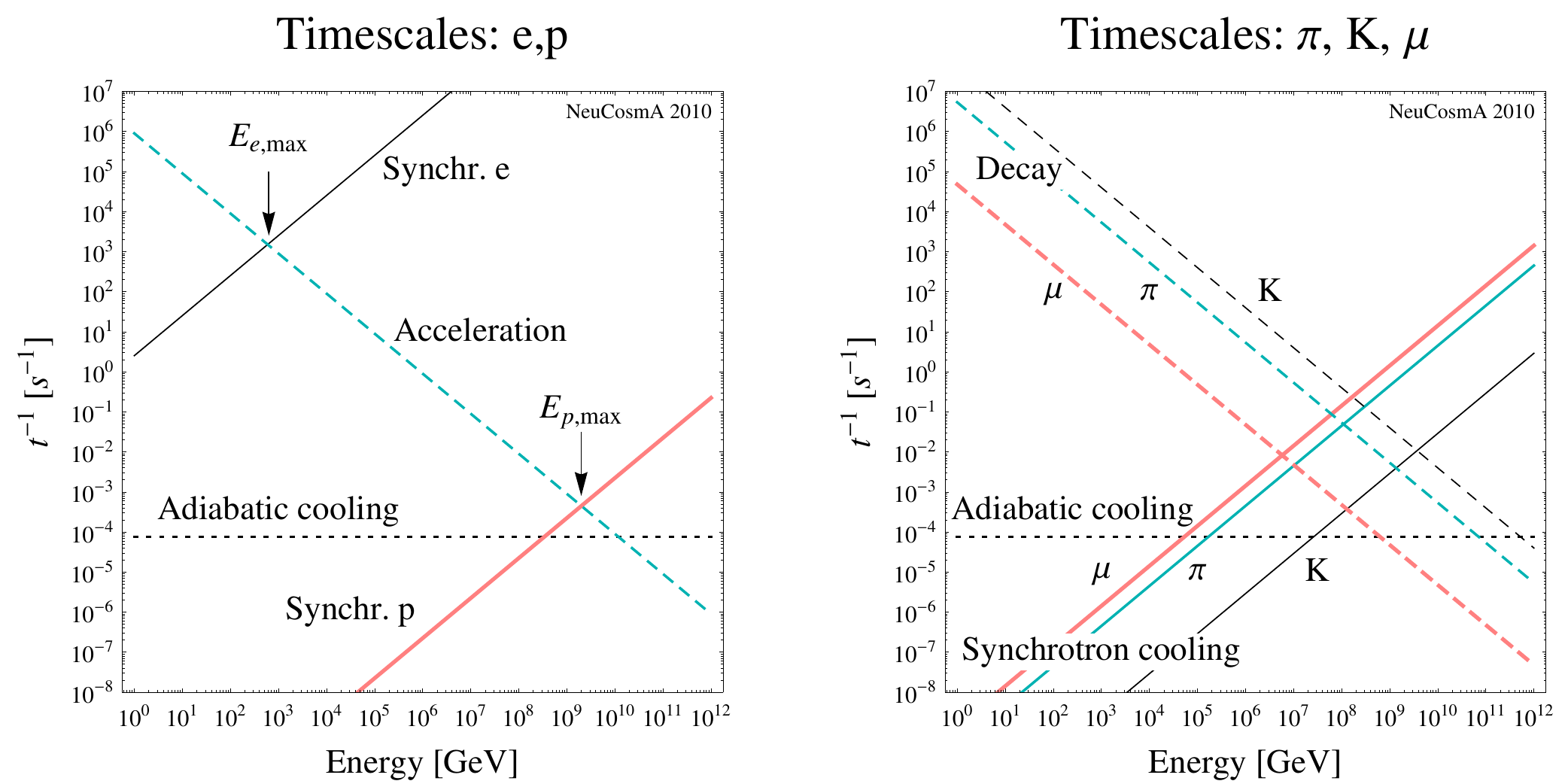}
\mycaption{\label{fig:timescales} Inverse timescales for electrons and pions in the left panel and for pions, kaons and muons in the right panel. Here the parameters  $R=10^{9.6}\,\mathrm{km}$, $B=10^3\,\mathrm{G}$,  $\alpha=2$, $E_{p,\mathrm{min}}=m_p$, $E_{e,\mathrm{min}}=m_e$, and $\eta=0.1$ have been used.}
\end{figure}

In the following, we discuss the results for the model with the parameters $B=10^3\,\mathrm{G}$, $R=10^{9.6}\,\mathrm{km}$, $\alpha=2$, $E_{p,\mathrm{min}}=m_p$, $E_{e,\mathrm{min}}=m_e$ and $\eta=0.1$ (which is going to be test point~3 in the next section). In Fig.~\ref{fig:timescales}, we show the different inverse timescales (rates) for electron and protons (left panel) and pions, kaons and muons (right panel) as a function of energy $E$. We can nicely see that the adiabatic cooling time is energy independent, the synchrotron loss rate proportional to $E$ and the acceleration and decay rate proportional to $1/E$. The dominant rate has always the largest values in Fig.~\ref{fig:timescales}. In the left panel we can read off the maximal energy for protons and electrons from the intersection point of the acceleration rate and the dominant energy loss rate (for comparison, see Fig.~\ref{fig:maxenergy}). In the right panel, the critical energies for muons, pions and kaons can be read off from the intersection point of synchrotron loss and decay rate. Taking into account the maximal energy of the protons, we can infer from the critical energies that energy losses are for this parameter set not important for kaons, but relevant  in the highest energy decade of the pion spectrum and the highest two energy decades of the muon spectrum. 

\begin{figure}[tp]
     \includegraphics[width=1\textwidth]{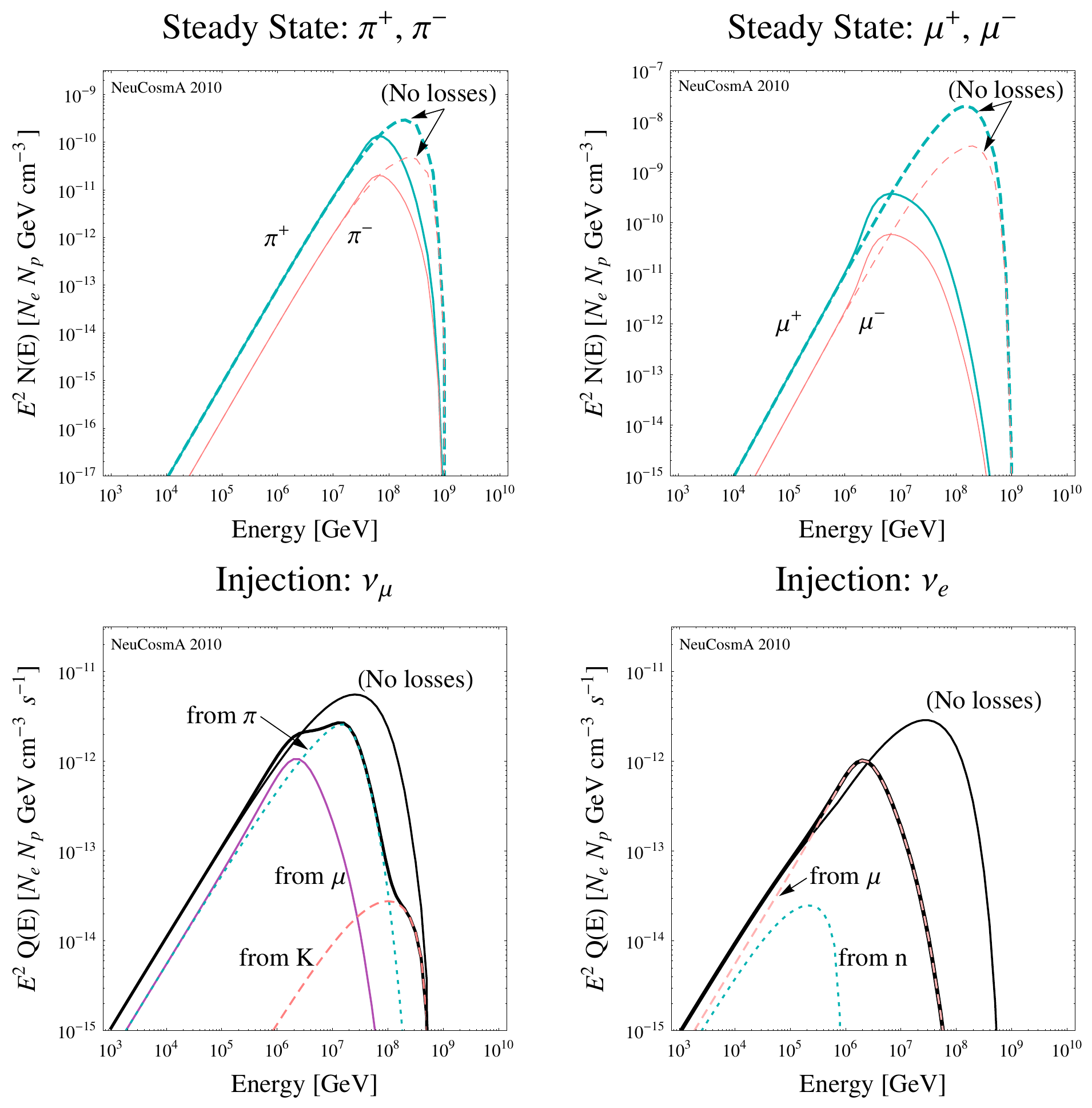}
\mycaption{\label{fig:spectraex} Pion (upper left), muon (upper right), muon neutrino (lower left), and electron neutrinos (lower right) spectra for the model parameters  $B=10^3\,\mathrm{G}$, $R=10^{9.6}\,\mathrm{km}$, $\alpha=2$, $E_{p,\mathrm{min}}=m_p$, $E_{e,\mathrm{min}}=m_e$, $\eta=0.1$ (neutrinos and antineutrinos added). In the upper panels, the thick solid curves represent the total spectra considering energy losses, and the dashed curves neglect energy losses of pions and muons. In the lower panels, the thin solid curves show the results without energy losses (summed over all decay modes).}
\end{figure}

 In Fig.~\ref{fig:spectraex}, we show in the upper row the steady state spectra for pions (left) and muons (right) considering energy losses (solid curves) and without energy losses (dashed curves). 
 Here the effect of energy losses sets in at around $10^8\,\mathrm{GeV}$ for pions and at $5\cdot10^6\,\mathrm{GeV}$ for muons, which are just the same values which can be read off from  Fig.~\ref{fig:timescales} (right panel). The muon energy spectrum changes the spectral index at both of these energies, because the parent pions are cooled above $10^8\,\mathrm{GeV}$. The qualitatively predicted behavior of the spectral indices leading to \equ{coolana} can be clearly seen. In addition, there is a small ``pile-up'' effect in the muon spectra at around $10^6\,\mathrm{GeV}$, where the cooled muons from higher energies pile up to produce a flux higher than the uncooled one.

In the lower row of \figu{spectraex}, we illustrate the muon (left) and the electron (right) neutrino spectra coming from the different parents (\cf, \figu{flowchart}) comparing energy losses of the parents and no losses, where neutrinos and antineutrinos are added. For the muon neutrinos (left panel), one can clearly see that  if there are large enough magnetic field effects, the different mass dependent cooling rates will lead to a spectral split of the corresponding neutrino spectra. The electron antineutrino spectrum from neutron decays is in this case sub-leading. It only contributes at low energies. In this specific example, the qualitative discussion above reproduces the spectral index of the neutrinos $\alpha_\nu=1$ below the cooling breaks, and the change of the spectra index at the breaks very well. However, the second break in the spectrum from muon decays is hardly visible.

In the following we will, unless noted otherwise, fix some of  the parameters ($\eta=0.1$, $E_{p,\mathrm{min}}=m_p$ and $E_{e,\mathrm{min}}=m_e$), since they do not exhibit the most interesting effects. If one changes the acceleration efficiency, one changes the maximal energy of the protons and electrons and subsequently of all other particles from the decay chain. If one changes the minimal energy of the protons in the injection spectrum, the steady state spectrum has  an index $\alpha^s_p=1$ below the minimal energy compared to $\alpha^s_p=2$ above (if adiabatic energy losses dominate). This break translates into the pion and subsequently into the neutrino spectrum. In order to observe this effect in the important region of the neutrino spectrum, the $E_{p,\mathrm{min}}$ has to be chosen higher than  about three orders of magnitude below the maximal proton energy, in our example, above $10^6 \, \mathrm{GeV}$ corresponding to a pre-acceleration $\gamma_p \sim 10^6$, which is rather high. On the other hand, a larger value of $E_{e,\mathrm{min}}$ 
translates into a break in the steady state spectrum with a spectral index $\alpha^s_e=2$ below and $\alpha^s_e=3$ above. This break translates into a break in the photon spectrum around $\varepsilon_c(E_{e,\mathrm{min}})$, see \equ{singlesyn}, dependent on the magnetic field.
The additional break will translate into the pion and neutrino spectrum when $E_p\,\varepsilon_c(E_{e,\mathrm{min}})$ (with $E_p\,<\,E_{p,\mathrm{max}}$) is above the threshold for photohadronic interactions. The effect is then visible in the upper part of the pion and neutrino spectra. It turns out that the break in the neutrino spectrum has to be about three orders of magnitude below the maximal neutrino energy to be significant. For the model parameters chosen in this section, this translates into $E_{e,\mathrm{min}} \gtrsim 1\,\mathrm{GeV}$, corresponding to a pre-acceleration $\gamma_e \sim 2000$. 
Of course, the pre-acceleration $E_{e,\mathrm{min}}$ and $E_{p,\mathrm{min}}$ depends on the specific model. For example, 
in some models a pre-boost comparable to the bulk Lorentz factor $\Gamma$ may be a reasonable assumption. Since the effects on our results are, however, small and of secondary interest, we choose relatively low minimal injection energies. Note, however, that one could use a relatively high $E_{e,\mathrm{min}}$ to model a (weak) break in a GRB photon field.

\subsection{Limitations of the model}
\label{sec:limitations}

Our model is based on the simplest plausible set of assumptions, which means that it naturally cannot describe arbitrary astrophysical sources. For example, consider the target photon field. Since it is produced by the electrons with the same spectral index as  the protons, the spectral index of the photon field is already fixed. Consider the often assumed photon field of GRBs with the spectral indices of $1$ (below the break) and $2$ (above the break), which is, in general, difficult to model with synchrotron radiation. In our model,  $\alpha_\gamma=1$ would require a hypothetical injection index for protons and electrons of $\alpha=0$, which again leads to a neutrino index of $\alpha_\nu=0$ and not the typical $\alpha_\nu=2$ (see \Sec~\ref{sec:typresults}). To obtain the index $\alpha_\nu=2$ for the neutrino spectrum, we have to choose $\alpha=4$, but then the index of the photon spectrum is $\alpha_\gamma=3$. We see that arbitrary combinations of indices of proton and photon spectrum are not possible since the injection index of electrons and protons is assumed to be equal and the photon index directly follows from the electron index. However, as mentioned above, one could use $E_{e,\mathrm{min}}$ to model at least some spectral break. This problem arises mainly for GRBs and not  for AGNs, since their photon fields can be described by synchrotron radiation much better. In that sense, our model is more a AGN-like model. However, to study the effects of secondary cooling, we use $\alpha=4$ in some cases, which may be unrealistically soft, but exhibits the features GRBs with respect to the pion spectra.

In addition, with our model, we can not describe astrophysical environments where processes such as inverse Compton and Bethe Heitler pair production (with high radiation densities) are dominant or at least equivalent to the considered energy losses. However, in these cases, the assumption of optically thin sources to neutrons may also no longer be reasonable since the interaction rate of neutrons with photons depends on the photon field density. In high radiation densities, the interaction rate may be higher than the escape rate, which means that the neutrons would interact (and also cool via photohadronics) before leaving the source. For example, the synchrotron photon density is proportional to the absolute normalization of the electrons and the size of the source. Hence, in case of high radiation densities, one has to consider the normalizations of the injection spectra separately, or the ratio between protons and electrons, depending on the choice of parameters. 

In the calculation of the photon field we assume that we are in the slow cooling regime. This means, that the dynamical timescale of the system is shorter than the cooling timescale. In the fast cooling regime, the dynamical timescale is longer than the radiative cooling timescale, as it may apply to the prompt emission of GRBs. The resulting photon spectrum is then softer by a factor of $1/2$. The index of the photon spectrum is $\alpha_\gamma^\mathrm{fast}=\alpha_e^s/2+1$ instead of $\alpha_\gamma^\mathrm{slow}=(\alpha_e^s-1)/2+1$ (see \cite{Piran:2004ba}).
Note that we also neglect neutrino production through pp interactions. This process may contribute if high energetic protons run into cold protons, \eg, in GRBs during the afterglow. Including it into our model, however, would require more parameters.

Finally, in our model we are exclusively interested in the target photon spectra for photohadronic interactions and not in the photon spectra observed at the Earth. To calculate the observed photon spectra, one has to take into account all radiation processes such as cascade radiation, synchrotron self absorption, and inverse Compton processes. However, note that this also implies that the observed photon field may not be the target photon field to be considered for photohadronic processes. Since our model is self-consistent, it does not rely on specific assumptions with respect to the multi-messenger connection.

\section{Neutrino fluxes and flavor ratios at the source}
\label{sec:source}

Here we discuss the neutrino fluxes, flavor ratios, and (electron) neutrino-antineutrino ratios at the source. In particular, we discuss the classification of our sources as a function of $R$ and $B$, as used in the Hillas plot. 

\subsection{Source classification and the Hillas plot}

\begin{figure}[tp]
\centering
\includegraphics[width=\textwidth]{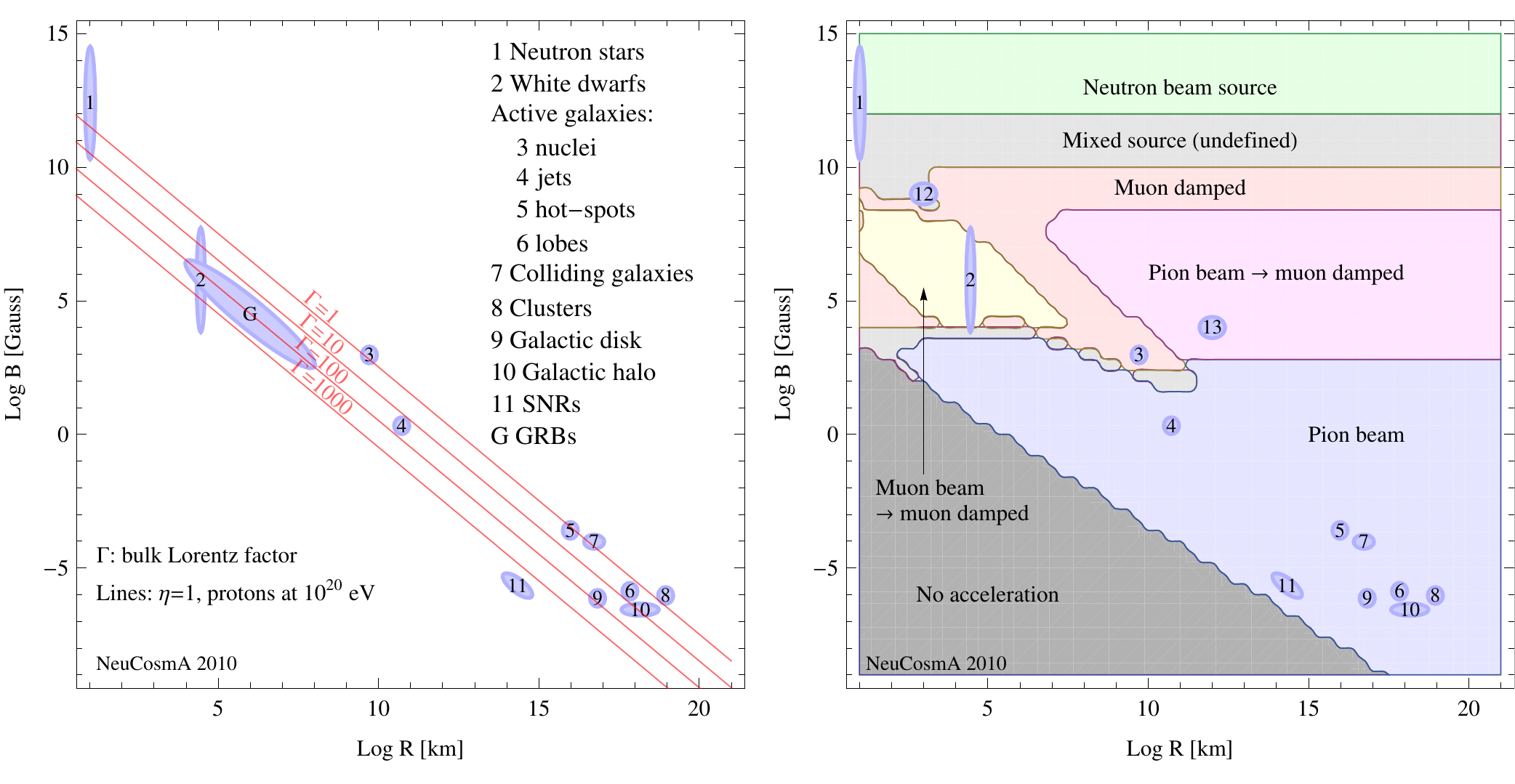}
\mycaption{\label{fig:hillas} Left panel: Possible acceleration sites in Hillas plot as a function of $R$ and $B$ (version adopted from M. Boratav). Right panel: Classification of sources for injection index $\alpha=2$ in this plot (see main text). Some points from left plot are shown for orientation, as well as two new points (12 and 13) are defined for later reference. }
\end{figure}

From \equ{hillas}, we can read off that if protons are accelerated ($Z=1$) and the acceleration efficiency is high $\eta \simeq 1$, the observation of the highest energetic cosmic rays with $E \simeq 10^{11} \, \mathrm{GeV}$ leads to the necessary condition
\begin{equation} 
B \, [\mathrm{G}] \gtrsim 1/3 \cdot 10^{13} (R \, [\mathrm{km}])^{-1}
\label{equ:hspec}
\end{equation}
 to produce such high energetic cosmic rays. Although we do not necessarily have to meet this condition for efficient neutrino production, it may serve as a hint for interesting source candidates. 
We show possible acceleration sites in the Hillas plot as a function of $R$ and $B$ in \figu{hillas}, left panel (version adopted from M. Boratav; \cf, \Ref~\cite{Ptitsyna:2008zs} for a more detailed recent discussion). One complication in this type of figure is that $R$ and $B$ using the argument in \equ{hillas} are potentially given in the SRF, whereas $E_{\mathrm{max}}$ is to be determined in the observer's frame. If the particles are accelerated in a relativistically moving environment, such as in a GRB fireball, this assignment is not trivial anymore, and the Lorentz boost $\Gamma$ of the acceleration region must be taken into account by boosting $R$ and $B$. We stick to the interpretation of $R$ and $B$ in the SRF, which means that (for $\eta=1$) the condition in \equ{hspec} depends on the Lorentz boost of the source. This is illustrated by showing \equ{hspec} for several selected Lorentz boosts in \figu{hillas}, left panel. We also include a GRB region ``G'' for illustration.\footnote{GRB prompt emission; for typical parameters, which can be estimated from empirical relations, see, \eg, \Refs~\cite{Guetta:2003wi,Piran:2005qu}. Note that the size of the acceleration region can be estimated from the timescale of the fluctuations in the GRB light curve (boosted into the SRF).}
 One can easily read off the figure that for the typical GRBs with $\Gamma \simeq 300$, the maximum energy in \equ{hspec} can be reached, since the line $\Gamma=100$ is partially exceeded. Note that some of the other regions also involve moderate boosts, which means that the interpretation in terms of $R$ and $B$ in the SRF has to be done carefully. For us, this interpretation is irrelevant, since we will use some points marked in \figu{hillas} as test points with $R$ and $B$ given in the SRF, and we leave the interpretation to the reader. In addition, note that the condition in \equ{hspec} is not necessarily matched by our model. Too large magnetic fields will lead to proton cooling by synchrotron radiation (such as for test points~1 and~2), which means that the maximum energy will be constrained by the synchrotron loss time in our model (\cf, \figu{maxenergy}). In addition, we use $\eta \simeq 0.1$ as acceleration efficiency in the following.

The discussion at the source implies that all quantities are given in the SRF, such as magnetic field $B$ and extension of the acceleration region $R$. In addition, the neutrino spectra are given at the source in arbitrary units (per energy, volume, and time) without Lorentz boost,  redshift, or flavor mixing.  This discussion has the advantage that very few source specific parameters have to be taken into account.  However, detector specifics, such as the detection threshold, cannot be taken into account for a source with arbitrary boost factor. For example, even if the neutrino energies are below the threshold at the source, the actual detectability depends on the boost factor, redshift, detection threshold, luminosity, and backgrounds. 

In the literature, typically the following three classes of neutrino sources are considered:
\begin{description}
\item[Pion beams] Neutrinos are produced by charged pion decays in the ratio $\nu_e$:$\nu_\mu$:$\nu_\tau$ of 1:2:0 (neutrinos and antineutrinos added). This means that the electron to muon neutrino ratio $R_e = (Q_{\nu_e}+Q_{\bar \nu_e})/(Q_{\nu_\mu}+Q_{\bar \nu_\mu}) \simeq 1/2$. In fact, corrections from the helicity dependence of the muon decays lead to small deviations from this ratio, which, however, can be understood in terms of particle physics, as well as there can be pile-up effects~\cite{Lipari:2007su}. The deviations depend on the input particle spectra (see, \eg, \Ref~\cite{Hummer:2010vx}) and are below the level of 10\%  in $R_e$. They are fully taken into account in our computations.
\item[Muon damped sources] If the synchrotron loss time scale of the muons is shorter than the decay time scale, the muons loose energy before they decay. In this case, only neutrinos from pion decays are present, leading to 
$\nu_e$:$\nu_\mu$:$\nu_\tau$ of 0:1:0 (neutrinos and antineutrinos added), or $R_e = (Q_{\nu_e}+Q_{\bar \nu_e})/(Q_{\nu_\mu}+Q_{\bar \nu_\mu}) \simeq 0$.
\item[Neutron beam sources] Given the assumption of sources optically thin to neutrons, the neutrons leave the source and decay into protons, electrons, and electron antineutrinos. These neutrinos are typically found at low energies, unless the synchrotron losses of the pions and muons are so large (\ie, the magnetic fields are very large) that the neutrinos from neutron decays contribute significantly. In this case, we have  
$\nu_e$:$\nu_\mu$:$\nu_\tau$ of 1:0:0 (neutrinos and antineutrinos added), or $R_e = (Q_{\nu_e}+Q_{\bar \nu_e})/(Q_{\nu_\mu}+Q_{\bar \nu_\mu}) \rightarrow \infty$. Note that there are other potential sources of neutrons, such as neutrons produced by the photo-dissociation of heavy nuclei~\cite{Anchordoqui:2003vc,Hooper:2004xr}.
\end{description}
In addition, we find a new class of sources:
\begin{description}
\item[Muon beams]  If the synchrotron loss time scale of the muons is shorter than the decay time scale, the muons loose energy before they decay. These muons may pile up at lower energies, where, similar to a neutrino factory, only neutrinos from muon decays are present. This leads to $\nu_e$:$\nu_\mu$:$\nu_\tau$ of 1:1:0 (neutrinos and antineutrinos added), or $R_e = (Q_{\nu_e}+Q_{\bar \nu_e})/(Q_{\nu_\mu}+Q_{\bar \nu_\mu}) \simeq 1$. Typically, a muon beam comes together with a muon damped sources at higher energies: The muons missing at high energies are recovered at lower energies. Note that there may be other sources of muon beams, such as heavy flavor decay ($D$, $D_S$, $B$, \etc) dominated sources, where the pions and kaons interact before they can decay, whereas the heavy mesons have shorter lifetimes (see, \eg, \Ref~\cite{Pakvasa:2010jj} for a discussion).
\end{description}

The above classification of sources in the literature is typically performed without considering the energy dependence explicitely. In order to consider the full energy dependence, we have to take into account the energy ranges where most neutrinos will be detected, \ie, the energy ranges around the spectral peaks.  The neutrino effective area roughly scales as $E^2 Q$ (for muon tracks) below about 10~TeV even in the presence of Earth attenuation, as it can be seen in Fig.~6 of \Ref~\cite{GonzalezGarcia:2009jc}. Therefore, we compute $E^2 (Q_{\nu_e}+Q_{\bar \nu_e})$ and $E^2 (Q_{\nu_\mu}+Q_{\bar \nu_\mu})$ separately, and then we determine the energy ranges where these fluxes are at least 1\% of the respective maximum. The relevant energy range used for the classification of the source is then the union of these ranges. As we have explained above, we do not take into account the boost at this level. However, neutrino energies below 0.1~GeV at the source cannot contribute to the event rate, since these will be below the typical muon track thresholds (100~GeV for IceCube) even for extremely large $\Gamma \simeq 1000$. Therefore, we do not consider such low energies. Within the identified energy ranges, we then classify according to: If $\nu_e$:$\nu_\mu$:$\nu_\tau$ is between $0.9$:$0.1$:$0$ and $1$:$0$:$0$, \ie, $R_e>9$, we have a neutron beam source.  If $\nu_e$:$\nu_\mu$:$\nu_\tau$ is between $0.1$:$0.9$:$0$ and $0$:$1$:$0$, \ie, $R_e<1/9$, we have a muon damped source source. If $R_e$ is between 0.45 and 0.55, we have a pion beam source. If $\nu_e$:$\nu_\mu$:$\nu_\tau$ is between $0.9$:$1.1$:$0$ and $1.1$:$0.9$:$0$, \ie, $0.9/1.1 \le R_e \le 1.1/0.9$, we have a muon beam source. 
Otherwise, the source is mixed (undefined). We require that $R_e$ is within the respective range over at least one order of magnitude in energy, otherwise the source is again undefined. Note that this classification as a function of energy also admits sources with several well defined energy regions. For example, we find sources which are pion beams at lower energies and then translate into muon damped sources at higher energies. 

We show the result of this classification for $\alpha=2$ (injection index) in \figu{hillas}, right panel, as a function of $R$ and $B$. Each of the points in this plot corresponds to a full model computation. The region in which no acceleration takes place or the photohadronic interactions are below threshold is marked by ``no acceleration''. Apart from that, all the characteristic sources mentioned above can be found. The classical pion beams cover most of the parameter regions, in particular, on the scales of galaxies and larger objects. Stronger magnetic fields lead to cooling, which means that the pion beams change into muon damped sources at higher energies (pion beam $\rightarrow$ muon damped). The muon damped region consists of two parts: For small $R \lesssim 10^7 \, \mathrm{km}$, the muons loosing energies typically pile up at lower energies, leading even to  separate ``Muon beam $\rightarrow$ muon damped'' sources as a function of energy. For larger $R$, the increasing magnetic field at some point leads to the domination of muon damping around the peak independent of $R$. For even higher magnetic fields, both pions and muons strongly loose energy, which means that the neutrinos from neutron decays enter the relevant energy range, which eventually leads to neutron beams.

\begin{sidewaysfigure}[tp]
\centering
\includegraphics[width=\textwidth]{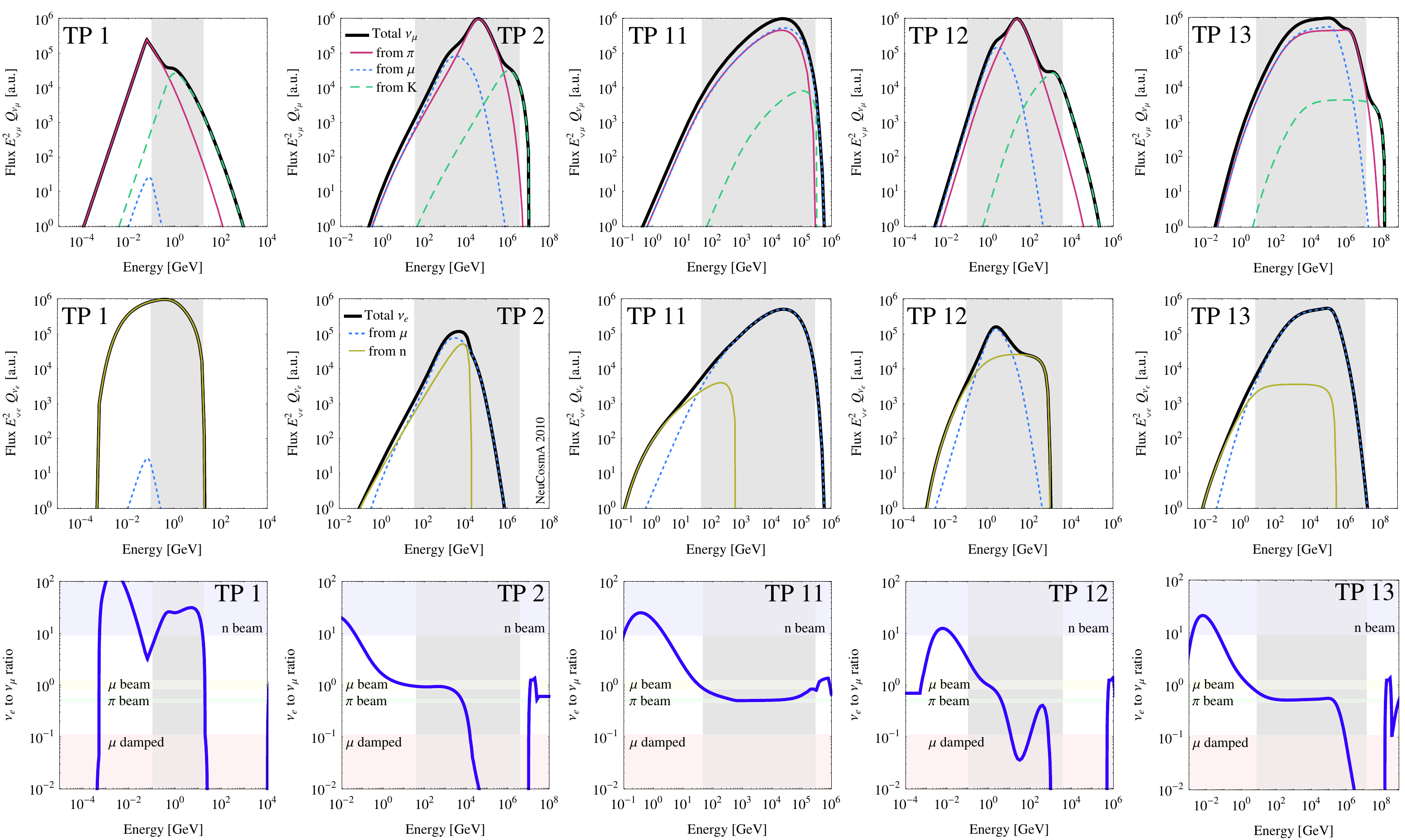}
\mycaption{\label{fig:spectra} Muon neutrino (upper row) and electron neutrino (middle row) spectra, as well as electron to muon neutrino flavor ratio (neutrinos and antineutrinos added, lower row) for several selected test points (in columns) from \figu{hillas}. The energy ranges used for the classification of the sources are gray-shaded. }
\end{sidewaysfigure}

In order to illustrate the classification of sources and typical examples even further, we show in \figu{spectra} the muon neutrino (upper row) and electron neutrino (middle row) spectra, as well as electron to muon neutrino flavor ratio (neutrinos and antineutrinos added, lower row) for several selected test points (in columns) from \figu{hillas}. In this figure, the energy ranges used for the source classification are gray-shaded according to the criteria above. Note that the strong jumps in the lower row above the energy classification windows may not be physical, since the fluxes are vanishing there. In the upper two rows, we show the total muon or electron neutrino spectra, as well as the individual contributions from pion, muon, kaon, and neutron decays. In the lower row, the ratio leading to the classification as neutron beam, pion beam, muon beam, or muon damped source is also shown. Test point~11 is a typical example for a pion beam, as it can be seen in the lower middle panel. Here neutrinos are mostly produced by pion and muon decays, whereas kaon and neutron decays are mostly outside the classification range. The main spectral index of the neutrino spectrum is roughly one, as expected for a proton spectrum not dominated by synchrotron losses. The neutrino energies are moderately high (in fact, test point~11 has the lowest maximal energies of the shown pion beam test points). If the magnetic field is increased, such as in test point~13, muon damping leads to a clear hierarchy among the neutrinos from kaon, pion, and muon decays in terms of their energy; cf, upper right panel. This source is classified as muon damped above about $10^5 \, \mathrm{GeV}$ (lower right panel). Over the whole energy range, it qualifies as ``pion beam $\rightarrow$ muon damped'' source, mainly because the spectrum is relatively flat for a wide range in $E^2 Q_\nu$. This flatness comes from the effect of proton synchrotron cooling. Compared to test point~13, test point~2 exhibits a much smaller acceleration region, which means that protons loose energy mostly by adiabatic cooling in our model. At lower energies, the muons damped from higher energies pile up, leading to the muon beam (\cf, second panel in third row). The spectra split between neutrinos from muon and pion decays can be nicely seen in the second panel of the first row.
 Test point~12 is an interesting example of a mixed source (see fourth panel in third row). In this case, even the pions and kaons are strongly cooled (see fourth panel in first row), whereas the neutrinos from neutron decays also show up in the relevant energy range (see fourth panel in second row).  The peak in the flavor ratio at about $10^2 \, \mathrm{GeV}$ corresponds to a narrow energy range where the neutron decays contribute, whereas for higher energies the muon neutrinos from kaon decays take over. Note that this effect from the kaon decays also leads to different conclusions than in \Ref~\cite{Moharana:2010su} in the GRB-relevant range, where it shadows the neutron decays at high energies (including neutral kaons, the effect may be even stronger). In addition, note that in test point~12 the maximum neutrino energies are relatively small, and even smaller in test point~1, where synchrotron cooling does not allow for high proton energies. Although test point~1 may be below the detection threshold of a typical neutrino telescope, it is an interesting example of a source where the neutron decays dominate (lower left panel) because all charged particle species loose energy very quickly in the extremely strong magnetic field. Unlike other neutron beams discussed in the literature, where neutrinos are produced by the photo-dissociation of heavy nulcei, we have a genuine neutron beam here, which does not depend on the properties of the interacting particles other than being charged. However, it should be noted that neutral kaon decays may have an effect here (which we do not consider), and that the source will probably not be detectable anyway because of the too low energies.  Test point~3, which lies close to three different regions (\cf, \figu{hillas}), will be discussed in the next section separately.

\subsection{Dependence on injection index}

\begin{figure}[tp]
\centering
\includegraphics[width=\textwidth]{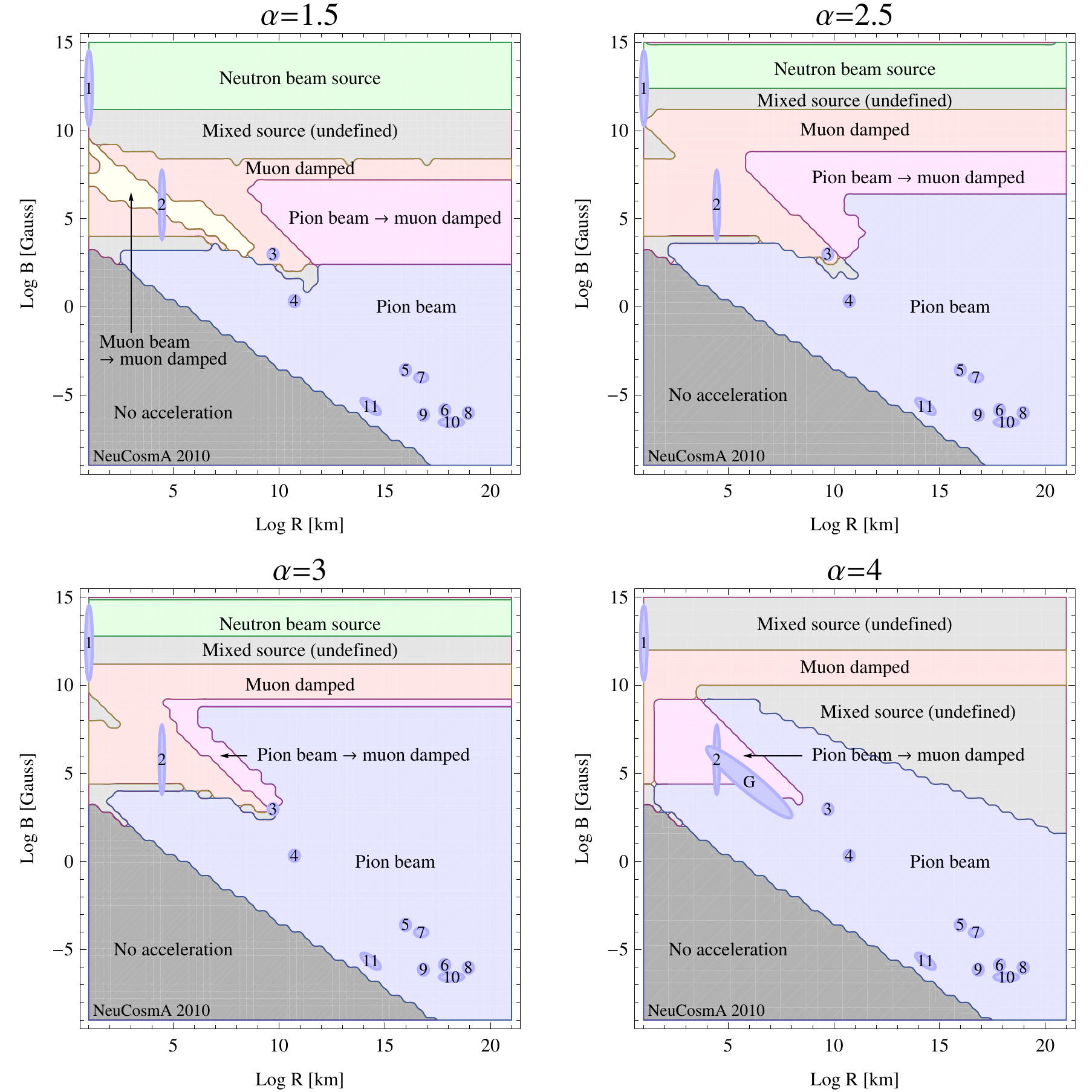}
\mycaption{\label{fig:hillasall} Classification of sources in the Hillas plot for different injection indices as shown in the plot labels. }
\end{figure}

Let us now discuss the dependence of the source classification on the universal injection index $\alpha$. While $\alpha=2$ may represent the often used standard case, many sources may have softer injection spectra, \ie, $\alpha > 2$. Therefore, we show in \figu{hillasall} the source classification for four different values of $\alpha$ other than $\alpha=2$ in \figu{hillas}: $\alpha=1.5$, $\alpha=2.5$, $\alpha=3$, and $\alpha=4$. While the case $\alpha=4$ seems to be unrealistically soft at first, the pion and muon neutrino spectra  in our model come then closest to the Waxman-Bahcall flux for GRBs~\cite{Waxman:1998yy} in the plateau region. Therefore, it is not surprising that the GRB region (\cf, \figu{hillas}, left panel) lies within the ``pion beam $\rightarrow$ muon damped'' region, as it is demonstrated in \Ref~\cite{Kashti:2005qa}. 

\begin{figure}[tp]
\centering
\includegraphics[width=\textwidth]{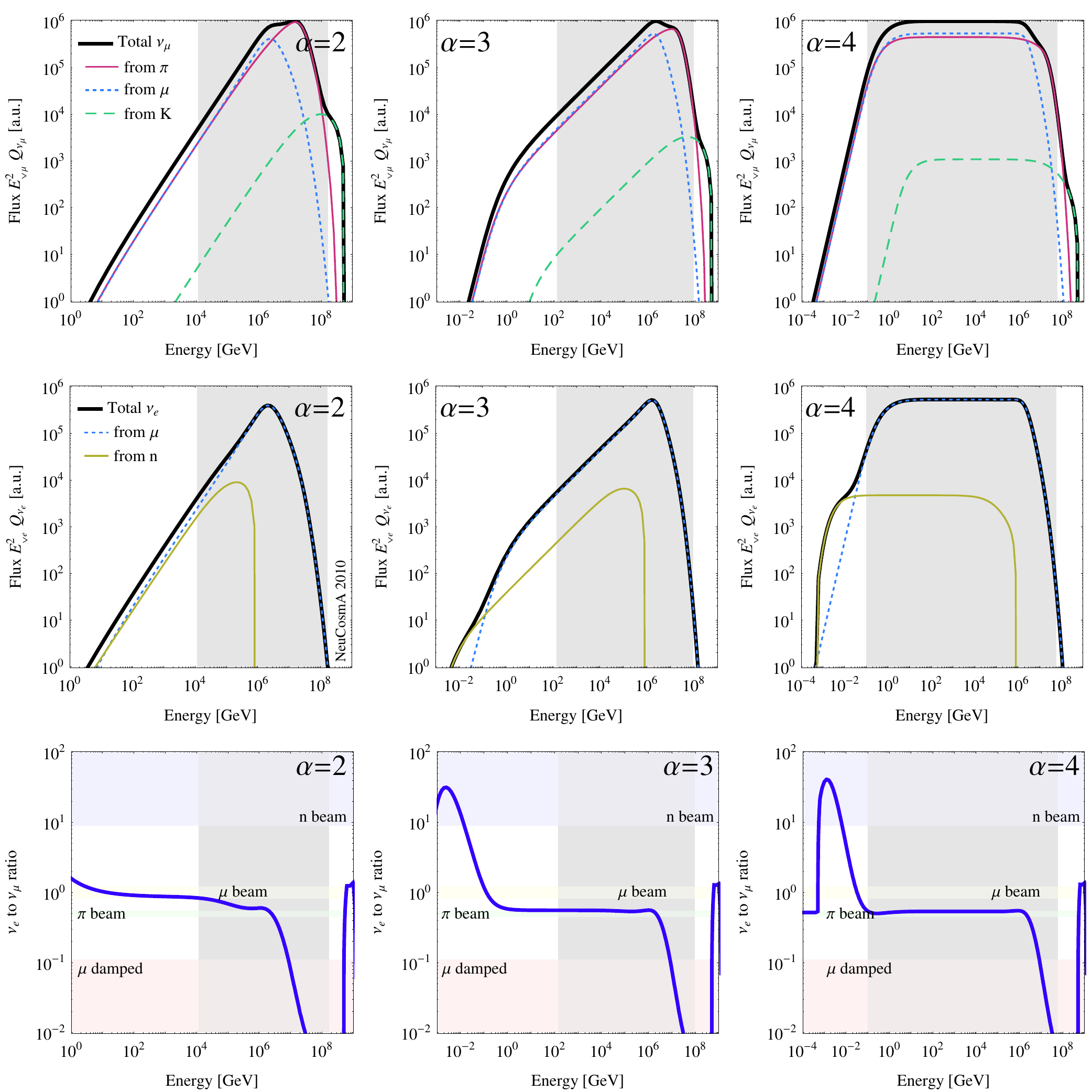}
\mycaption{\label{fig:spectratp3} Muon neutrino (upper row) and electron neutrino (middle row) spectra, as well as electron to muon neutrino flavor ratio (neutrinos and antineutrinos added, lower row) for test point~3 for different $\alpha$ (in columns.  The energy ranges used for the classification of the sources are gray-shaded. }
\end{figure}

While some regions are almost unaffected by the injection index, such as the one around test points~5 to~11, there is some transition in other parts of the parameter space. Let us first of all discuss test point~3, which lies close to three different regions for $\alpha=2$ (\cf, \figu{hillas}), and for which the classification changes as a function of $\alpha$. We show the neutrino spectra and flavor ratios for test point~3 and three different injection indices in \figu{spectratp3}. For $\alpha=2$, we have a muon damped source, because the lower energy range in the classification window exceeds the pion beam allowed range (see lower left panel). The reason is a combination of different factors, such as a small pile-up effect, the kinematics of the weak decays, and the non-negligible contribution from neutron decays (see middle left panel).  For $\alpha=3$, the source qualifies as pion beam for most of the energy range. In this case, the contribution from neutron decays is relatively smaller.  At high energies, however, we have a muon damped source (see lower middle panel). For $\alpha=4$, the muon damped range becomes shorter than the required order of magnitude (see lower right panel). Here the spectrum is flat in $E^2 Q_\nu$, which means that the statistics will be dominated by a wide energy range, and the muon damped part only gives a minor contribution.

Apart from test point~3, there are a number of different interesting regions in \figu{hillasall} where the classification changes as a function of $\alpha$. For example, consider test point~2, which is classified as ``pion beam $\rightarrow$ muon damped'' for $\alpha=4$. For smaller $\alpha$, however, the spectral shape changes, and, in the most extreme case $\alpha=1.5$, the neutrinos from muon and pion decays lead to two separated sharp peaks. In between these two extremes, the above mentioned combination of factors (as for test point~3) prevents the identification of the pion beam part, and leads to a muon damped source. On the upper end of the plot (large $B$), the dominance of neutron decays in the classification range depends on the injection index. For $\alpha > 2$, the neutron beam region is pushed towards higher $B$ until it vanishes, because then kaon decays contribute significantly. Another interesting region is the one around $10^{13} \, \mathrm{km}$ and $10^7 \, \mathrm{G}$, which changes from ``pion beam $\rightarrow$ muon damped'' (small $\alpha \lesssim 2$) over ``pion beam'' (medium $\alpha \sim 3$) to an undefined source (large $\alpha \gtrsim 4$). Here the relative position of the peak of the spectrum (from the balance between injection and cooling) shifts to the left (to low energies) with increasing $\alpha$, such that first of all the muon damped part leaves the classification energy range. For even larger $\alpha$, strong pile-up effects and too low energies destroy the classification as pion beam.

\subsection{Optically thin $\boldsymbol{p\gamma}$ sources and the neutrino-antineutrino ratio}
\label{sec:pgsource}

Apart from the flavor ratios at the source, it is interesting to discuss which sources behave as optically thin (to neutrons) $p \gamma$ sources. In order to identify such sources, one would use the Glashow resonance process at the detector, which is only sensitive to $\bar \nu_e$.\footnote{Note that the Glashow resonance requires a particular neutrino energy, which, however, depends on the boost. Since we do not include a Lorentz boost yet, we classifiy the sources in the whole energy range in this section.} For an optically thin $p \gamma$ source, protons interact with photons leading to mainly $\pi^+$, which then decay into $\mu^+$ and finally into $\nu_e$ (\cf, \equ{piplusdec}). Therefore, almost no $\bar \nu_e$ are produced at the source. At the source, we therefore expect $Q_{\bar \nu_e}/Q_{\nu_e} \simeq 0$  for the ``optically thin $p\gamma$ source''. Note that if the source was {\em not} optically thin to neutrons, $n \gamma$ (and other) interactions would break the asymmetry between $\bar \nu_e$ and $\nu_e$. Neutron decays would also increase $Q_{\bar \nu_e}/Q_{\nu_e}$.
In addition, if the source produced neutrinos by $pp$ interactions, $\pi^+$ and $\pi^-$ would be produced in comparable amounts, which means that $Q_{\bar \nu_e}/Q_{\nu_e} \simeq 1$.  Therefore, we use the term ``optically thin $p \gamma$ source'' for $Q_{\bar \nu_e}/Q_{\nu_e} \simeq 0$, because the source has to be both optically thin to neutrons, and most of the neutrinos have to be produced by $p \gamma$ interactions followed by charged pion decays.

In practice, the above mentioned argument $Q_{\bar \nu_e}/Q_{\nu_e} \simeq 0$ only holds in the $\Delta$-resonance approximation \equ{ds} of the photohadronic interactions. If other processes, such as direct production and multi-pion production are included, the contribution from $\pi^-$ leads to some contamination from $\bar \nu_e$; see, \eg, \Refs~\cite{Mucke:1999yb,Hummer:2010vx}. Since we include these processes, we accept $Q_{\bar \nu_e}/Q_{\nu_e} \lesssim 0.2$ as optically thin $p \gamma$ source. We use the same energy range for this classification as above for the flavor ratios, and we require that $Q_{\bar \nu_e}/Q_{\nu_e} \lesssim 0.2$ over at least one order of magnitude in energy. Note that flavor mixing in combination with muon damping may introduce additional electron antineutrinos to the $Q_{\bar \nu_e}/Q_{\nu_e}$ ratio, which we will discuss later. Without flavor mixing, however, any produced electron antineutrinos will be strongly suppressed. In this case, the classification as optically thin $p \gamma$ source will be spoilt if neutron decays contribute significantly.

\begin{figure}[tp]
\centering
\includegraphics[width=\textwidth]{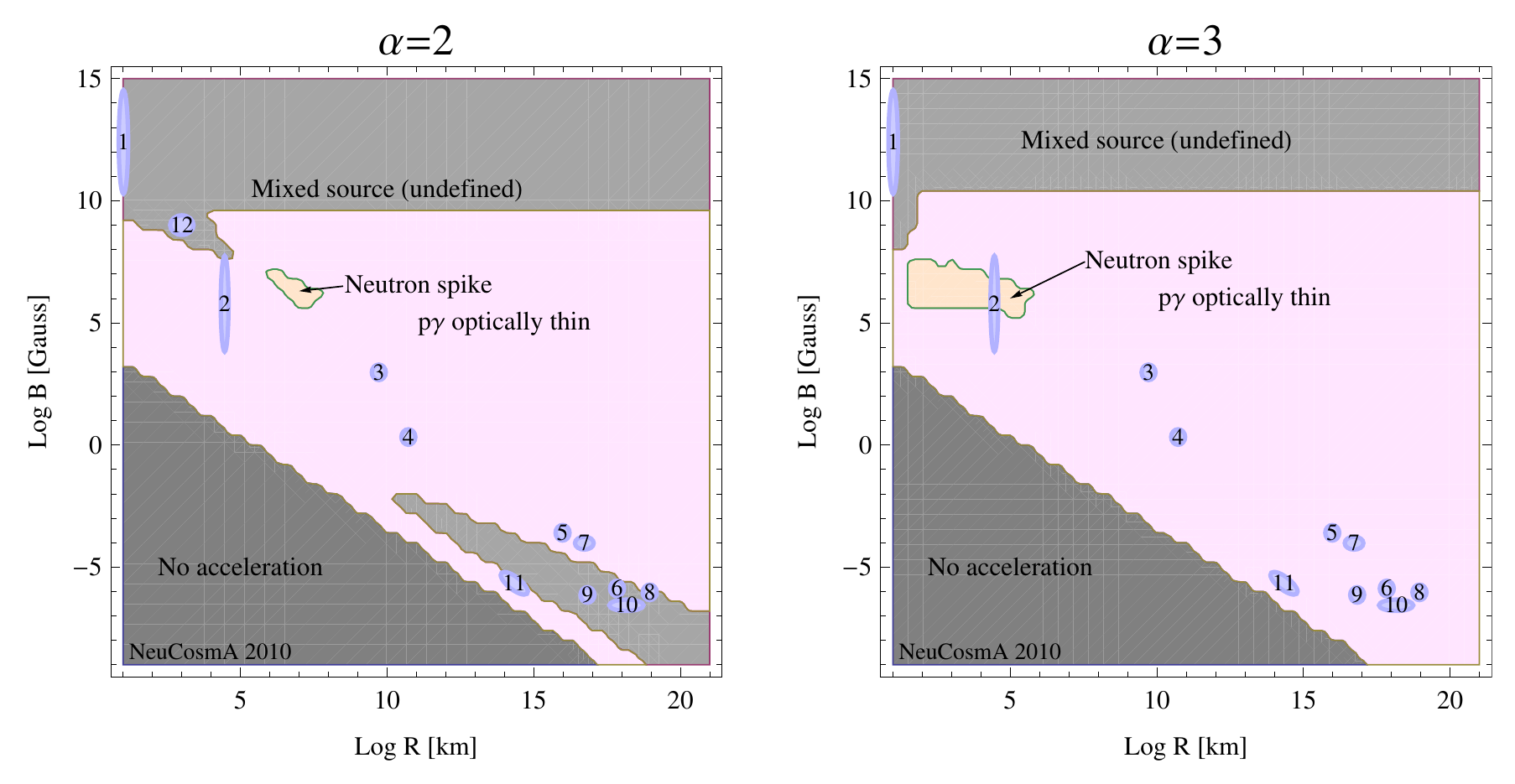}
\mycaption{\label{fig:hillaspgall} Classification of optically thin $p \gamma$ sources (see main text) in the Hillas plot for different injection indices as shown in the plot labels. }
\end{figure}

We show the region of optically thin $p\gamma$ sources for two different values of $\alpha$ in \figu{hillaspgall}. Apart from large $B$, where neutron decays significantly contribute to the spectrum, and a region for $\alpha=2$ around test points~6,~8,~9, and~10, most of the sources qualify as optically thin $p\gamma$ sources. As we demonstrate below, for large $B$, the contribution of neutron decays prohibits the classification as such source. The undefined source region in the lower right corner of the left panel, however, comes from the significant contribution of high energy events in the photohadronic interactions, which break the $\pi^+$-$\pi^-$-asymmetry.
There is only very little dependence on the injection index. However, this classification changes if flavor mixing is included within a slightly different analysis, as we will demonstrate later.

\begin{figure}[t!]
\centering
\includegraphics[width=\textwidth]{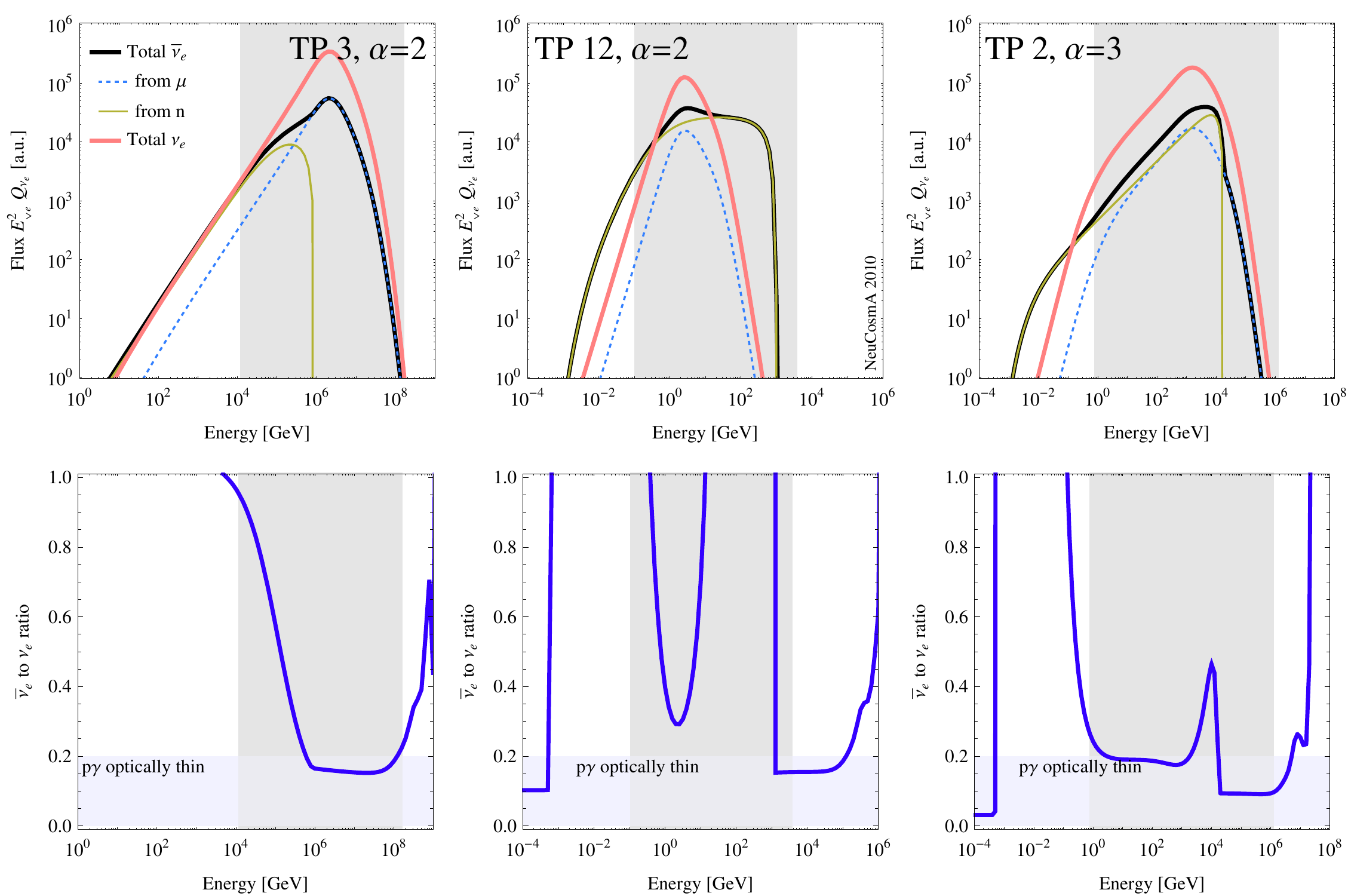}
\mycaption{\label{fig:spectrapg} Electron neutrino and antineutrino fluxes (upper row), as well as electron antineutrino to neutrino ratio (lower row) for several selected test points (in columns) from \figu{hillaspgall}.  The energy ranges used for the classification of the sources are gray-shaded. }
\end{figure}

We show several examples for selected test points and values of $\alpha$ in \figu{spectrapg}, where the $\nu_e$ and $\bar \nu_e$ spectra are shown separately. For example, in the left column (test point~3 for $\alpha=2$) we find a typical optically thin $p \gamma$ source within most of the classification energy range (gray-shaded). Here the neutrinos from neutron decays only dominant at the lower end of the classification range (see upper left panel). In the middle column, however,  neutron decays contribute significantly over the whole energy range, which means that the source cannot be classified as optically thin $p \gamma$ source. From the $\bar \nu_e$ to $\nu_e$ ratio, one can in this case not identify easily where the symmetry between electron neutrinos and antineutrinos actually comes from. The third column shows another example of a optically thin $p\gamma$ source. In this case, however,  neutron decays contribute significantly in a small energy window around $10^4 \, \mathrm{GeV}$, which leads to a ``neutron spike'' within the classification range. The regions where these neutron spikes occur are marked in \figu{hillaspgall}. They can be much more prononounced than for the shown test point.

\section{Neutrino flavor ratios at the detector}
\label{sec:detector}

In this section, we consider the neutrino fluxes at the surface of the Earth, including a possible boost of the acceleration region and neutrino flavor mixing. We consider neutrino flavor ratios as observables, which correspond to quantities in principle accessible to the detector. For these quantities, the absolute normalization cancels. We first show the impact of flavor mixing on the observables. Then we illustrate how flavor ratios can be used to extract information on the astrophysical parameters. And finally, we show the impact of Lorentz boost and flavor mixing on the Glashow resonance. Note that we do not include Earth attenuation effects in the present study.

\subsection{Impact of neutrino flavor mixing}

The easiest possibility to measure flavor ratios at the detector requires the identification of showers, which are much harder to detect than muons, and furthermore fiducial volume (not effective area) limited. Therefore, the statistics expected from showers is much smaller than from muon tracks.
If we assume that electromagnetic (from $\nu_e$) and hadronic (from $\nu_\tau$) showers do not need to be distinguished, a useful observable is the ratio between muon tracks and showers~\cite{Serpico:2005sz}
\begin{equation}
\hat R \equiv \frac{\phi_{ \nu_\mu}^{\mathrm{Det}}}{\phi_{\nu_e}^{\mathrm{Det}}+\phi_{\nu_\tau}^{\mathrm{Det}}}
\end{equation}
where neutrinos and antineutrinos of each flavor are summed over. 
Here $\phi_{\nu_\alpha}^{\mathrm{Det}}$ is the flux at the detector in units of $\mathrm{cm}^{-2} \, \mathrm{s}^{-1}$ $[\mathrm{sr}^{-1}]$, related to $Q_{\nu_{\alpha}}$ by a flavor independent normalization factor.
Note, however, that neutral current events will also produce showers, which, in practice, have to be included as background. The benefit of this ``flavor ratio'' is that the luminosity of the source drops out. In addition, it represents the experimental measurement with the simplest possible assumptions. At IceCube, showers cannot be expected to be identified below about 1~TeV, which we consider as threshold for this observable.

The neutrino flux at the detector is given by the neutrino flux per flavor $\phi_{ \nu_\alpha}^{\mathrm{Earth}}$  at the surface of the Earth (we do not consider Earth attenuation effects here), modified by the effects of flavor mixing 
\begin{equation}
\phi_{ \nu_\beta}^{\mathrm{Det}} = \sum\limits_{i=1}^3 \sum\limits_{\alpha} |U_{\alpha i}|^2 \, |U_{\beta i}|^2 \, \phi_{ \nu_\alpha}^{\mathrm{Earth}} \, . \label{equ:flmixing}
\end{equation}
In addition, we consider a possible Lorentz boost $\Gamma$ of the acceleration region, which leads to a Lorentz boost of the neutrino energy $E_\nu \rightarrow \Gamma \, E_\nu$.\footnote{Often, a more general Doppler factor is used, which includes the Lorentz boost and the viewing angle and possibly redshift effects. The value of the Doppler factor D corresponds to the Lorentz boost factor $\Gamma$ if the source is boosted at an angle $1/\Gamma$ relative to the observer (redshift ignored).} Note that the transformation from the neutrino flux at the source to the flux at the detector includes a number of normalization factors coming from Lorentz boost,  the integration over the production region, and the luminosity distance of the source. However, these will not affect the flavor ratios.

Since in \equ{flmixing} the neutrino mixing angles appear, the uncertainties of the mixing angles will lead to uncertainties in the flavor ratios. In order to discuss these, we use the following values and current $3 \sigma$ ranges (see \Ref~\cite{Schwetz:2008er}): $\sin^2 \theta_{23}=0.5$ ($3\sigma$: 0.36 ... 0.67), $\sin^2 \theta_{12}=0.318$ ($3 \sigma$: 0.27 ... 0.38), $\sin^2 \theta_{13}=0$ ($3\sigma$: $\sin^2 \theta_{13} \le 0.053$). In addition, we consider future improved bounds ($3\sigma$) on $\theta_{13}$ from Daya Bay $\sin^2 \theta_{13} \le 0.012$ and $\theta_{23}$ from T2K $0.426 \le \sin^2 \theta_{23} \le 0.574$, which represent the expected sensitivities at around 2015~\cite{Huber:2009cw}. Beyond these, a neutrino factory may improve the bounds even further: $\sin^2 \theta_{13} \lesssim 1.5 \, 10^{-5}$~\cite{Huber:2003ak} and $0.46 \le \sin^2 \theta_{23} \le 0.54$~\cite{Tang:2009na} ($3\sigma$), which we label ``2025''. For $\theta_{12}$, the errors may improve somewhat, but new generations of experiments are not assumed here. Since $\theta_{12}$ is of secondary importance in \equ{flmixing}, we do not consider such future improvements.

Let us now discuss how the source identification changes in the presence of flavor mixing. First of all, it does not make sense to consider an arbitrarily wide energy range, since the region close to the peak in $E^2 Q_\nu$ will contribute most significantly to the event rates. Therefore, we define, as in the previous section, the energy window which captures the upper two orders of magnitude in the flux as ``energy classification range''. However, we use the muon neutrino flux at the detector (neutrinos and antineutrinos added) for the determination of this window, since muon tracks are easiest to measure. Compared to the electron to muon neutrino ratio at the source (which may be very large for a neutron beam source), the observable $\hat R$ takes specific values for the considered source types and best-fit values: For the pion beam source, $\hat R = 0.5$, for the muon damped source, $\hat R \simeq 0.64$, for the muon beam source, $\hat R \simeq 0.44$, and for the neutron beam source $\hat R \simeq 0.28$. We mark these values in the figures where applicable.

\begin{figure}[tp]
\centering
\includegraphics[width=0.5\textwidth]{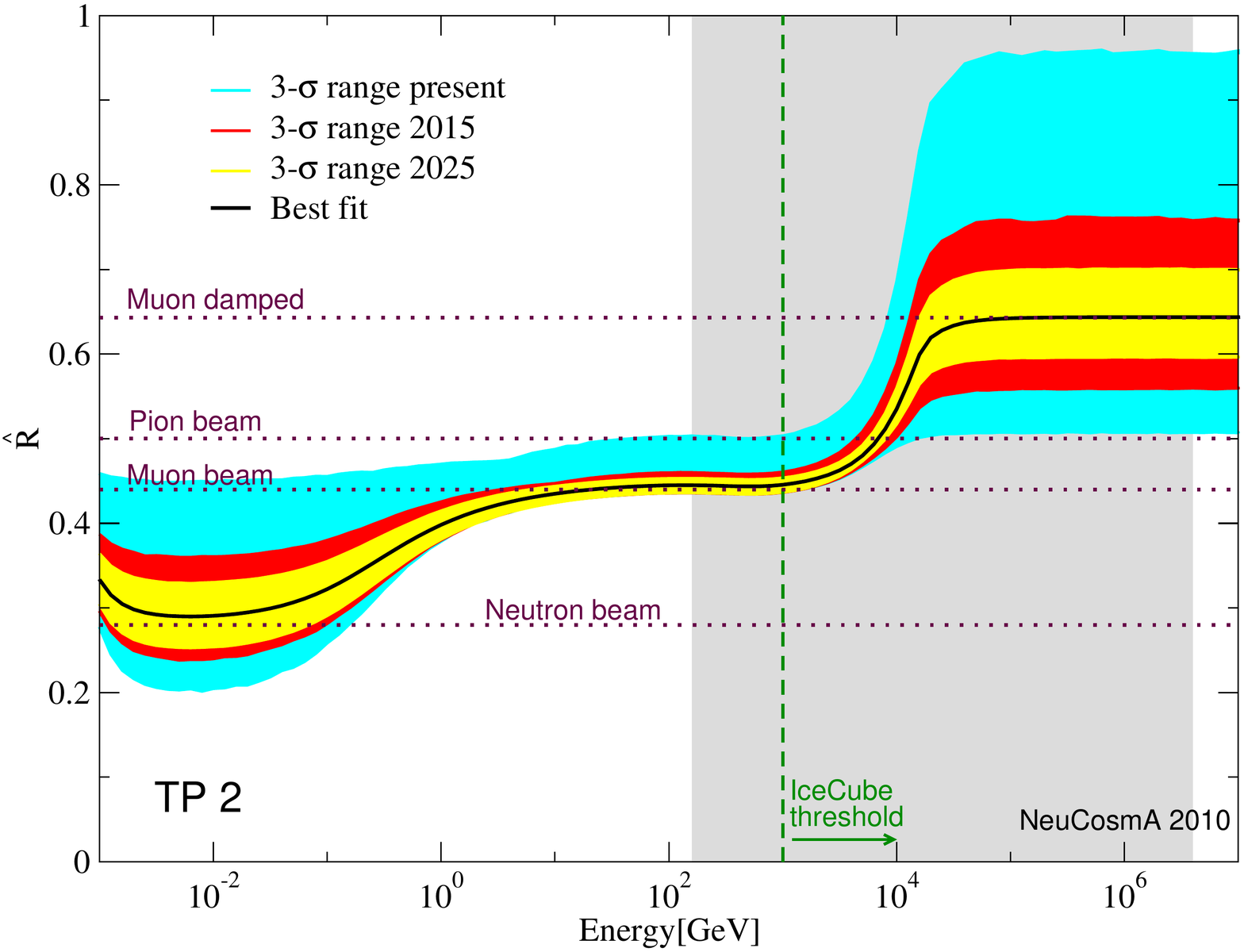} \\
\includegraphics[width=0.5\textwidth]{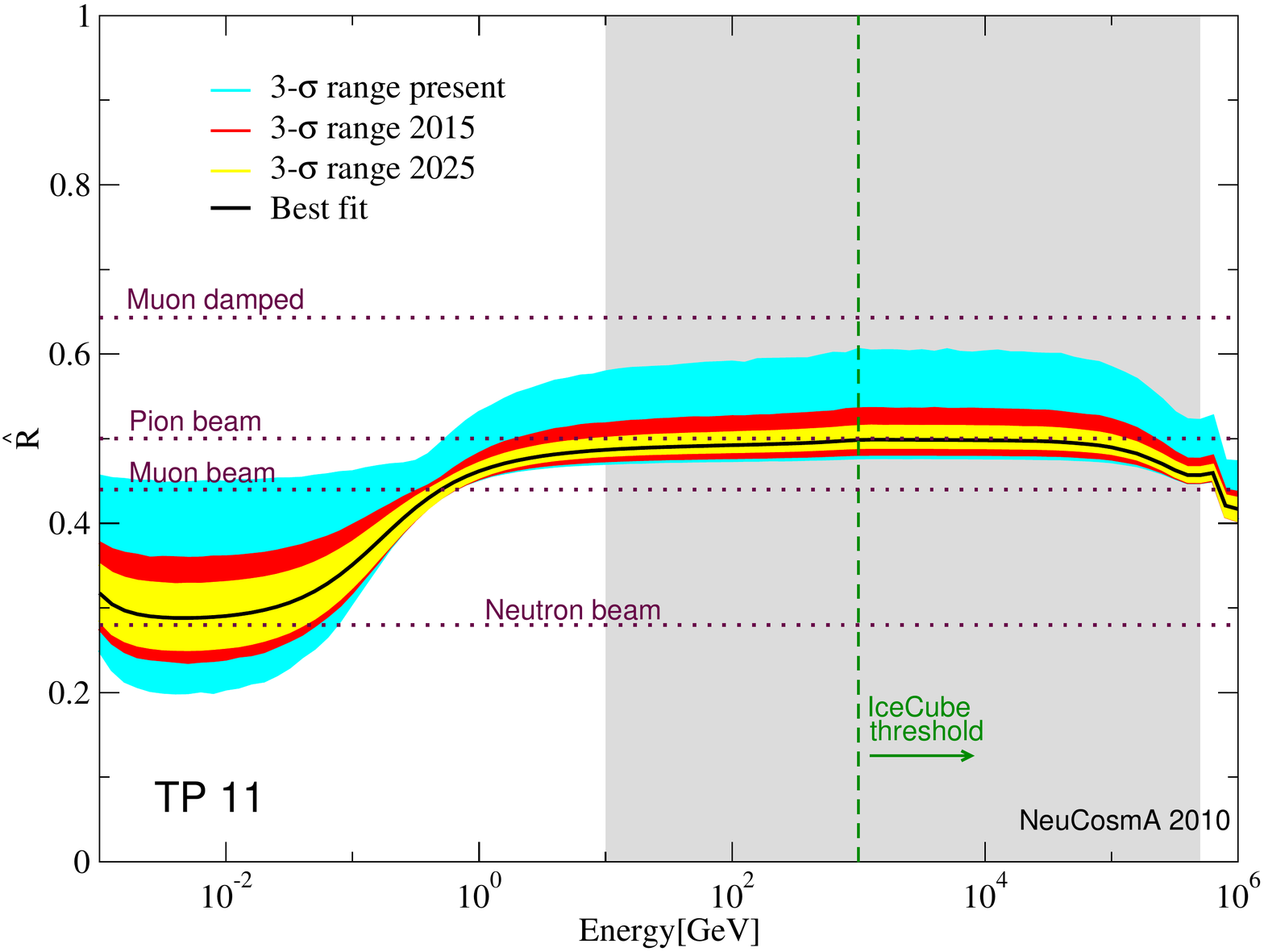} \\
\includegraphics[width=0.5\textwidth]{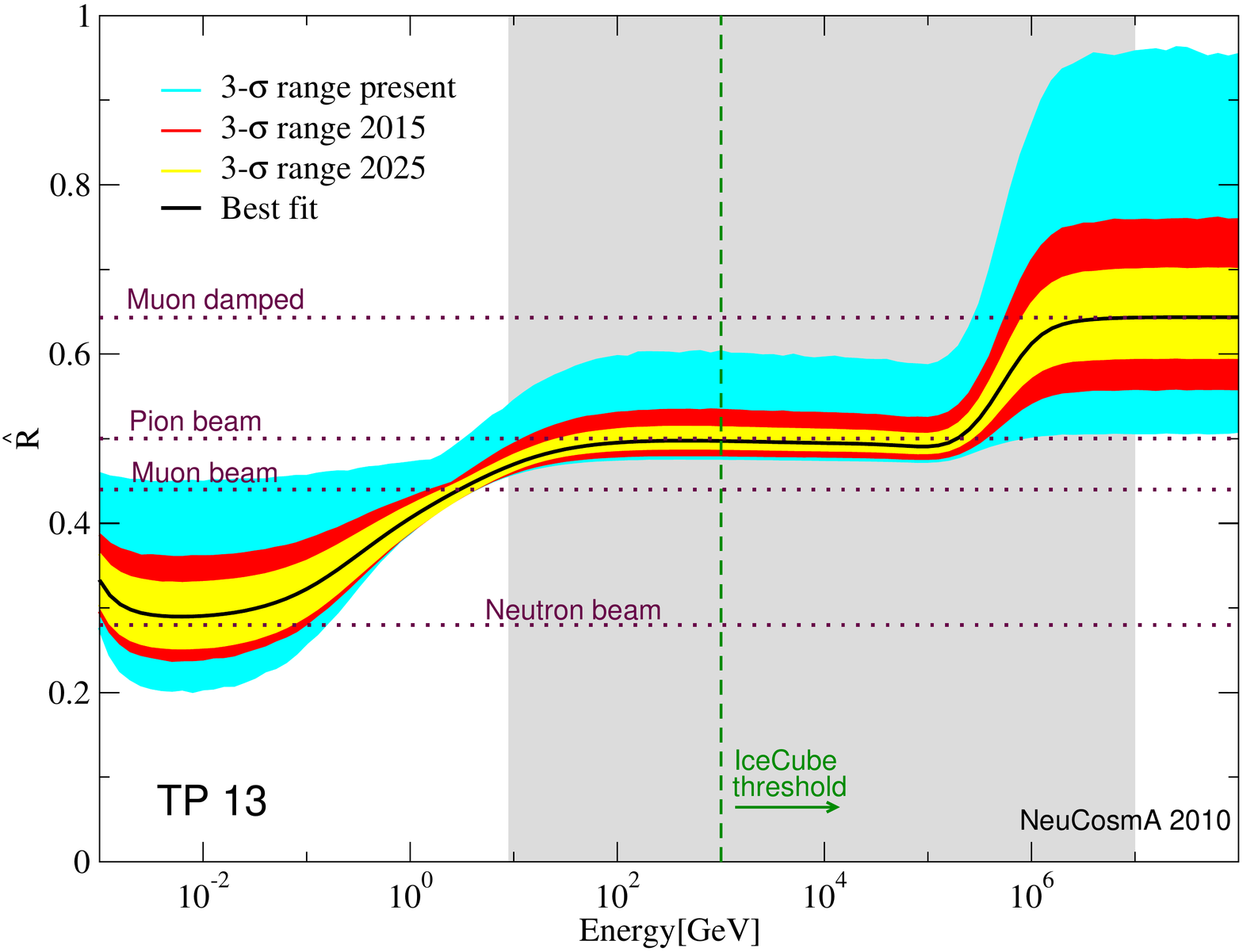}
\mycaption{\label{fig:flmix} Flavor ratio $\hat{R}$ after flavor mixing as a function of energy for test points~2, 11, and~13 from   \figu{hillas} (right panel). The solid curves show the result for the current best-fit values, whereas the shaded regions show the ranges if the neutrino mixing  parameters are varied within their current or future (two different assumptions) $3 \sigma$ ranges  (see main text for values). Shaded  rectangles mark the classification energy ranges where most events are expected (see main text). The flavor ratios for pion beams, muon damped sources, and neutron beams are marked. Lorentz boost factor  $\Gamma=1$ assumed. }
\end{figure}

We show in \figu{flmix} the flavor ratio $\hat{R}$ after flavor mixing as a function of energy for test points~2, 11, and~13 from  \figu{hillas} (right panel). The solid curves show the result for the current best-fit values, whereas the shaded regions show the ranges if the neutrino mixing  parameters are varied within their current or future (for 2015 and 2025) $3 \sigma$ ranges. Shaded  rectangles mark the classification energy ranges where most events are expected. In addition, the assumed IceCube threshold is shown. The flavor ratios for pion beams, muon damped sources, and neutron beams are marked. Here a Lorentz boost  $\Gamma=1$ is assumed, larger Lorentz factors would shift the spectrum (including classification range) to the right, but the IceCube threshold would remain fixed.

Comparing \figu{flmix} to \figu{spectra} for the current best-fit values (\cf, solid curves in \figu{flmix}), the energy dependence at the source translates into the energy dependence with the same features at the detector (only the vertical direction is inverted). Therefore, in principle, the source classes can be identified as a function of energy. The upper panel corresponds to a muon damped source, the middle panel to a pion beam source, and the lower panel to a pion beam $\rightarrow$ muon damped source. If, however, the current parameter uncertainties are included in the considerations, a particular value of $\hat R$ cannot be assigned to a specific source type anymore. For example, consider the lower panel: If $\hat R \simeq 0.5$, the source may be a pion beam, a muon damped source, or anything in between. As it can be read off from the figure, already the improved precision on the mixing parameters in about 2015 allows for a classification into the different source types. Even the muon beam range in the upper panel, $\hat R = 0.5$ (pion beam) could be, in principle, excluded. Note that we have chosen $3\sigma$ ranges in this figure, which is already quite conservative.
In addition, we have chosen $\theta_{13}=0$ as the best-fit, because otherwise we would need to discuss the measurement of $\deltacp$ as well. If Nature has chosen $\theta_{13}>0$, a neutrino factory may eventually be necessary to constrain $\theta_{13}$ and $\deltacp$ simultaneously.  The main uncertainty, however, comes from $\theta_{23}$.  A different value of $\Gamma$ would not change our results qualitatively, but instead shift a larger region beyond the assumed IceCube threshold.

\subsection{Measurement of astrophysical quantities}

Once neutrino flavor ratios are measured in neutrino telescopes, it is conceivable to use such observations to constrain the parameters of the source. We know, for instance, that the observation of supernova neutrinos is often used to constrain the physical processes that take place within the supernova, so the same  idea can be applied also to the neutrinos produced in cosmic accelerators.

Assume, for instance, that an optical (or gamma ray) counterpart observation of the source  has allowed the measurement of  $\Gamma$, $R$ and $\alpha$, \eg,  via the time dependence of the photon spectrum. Can $B$ be extracted from the energy dependence of the flavor ratio $\hat R$? In general, it is difficult to do so because $\hat R$ is not very sensitive to the magnetic field. In fact, we see from \figu{hillas} that, at a given $\hat R$, the magnetic field can generally vary over several orders of magnitude without changing the classification of the region. But,  close to a boundary between two regions, the magnetic field determines the classification of the source and, therefore, the expected value of $\hat R$. It is close to such boundaries that the measurement of $\hat R$ may allow to  constrain the magnetic field.

\begin{figure}[tp]
\centering
\includegraphics[scale=0.4]{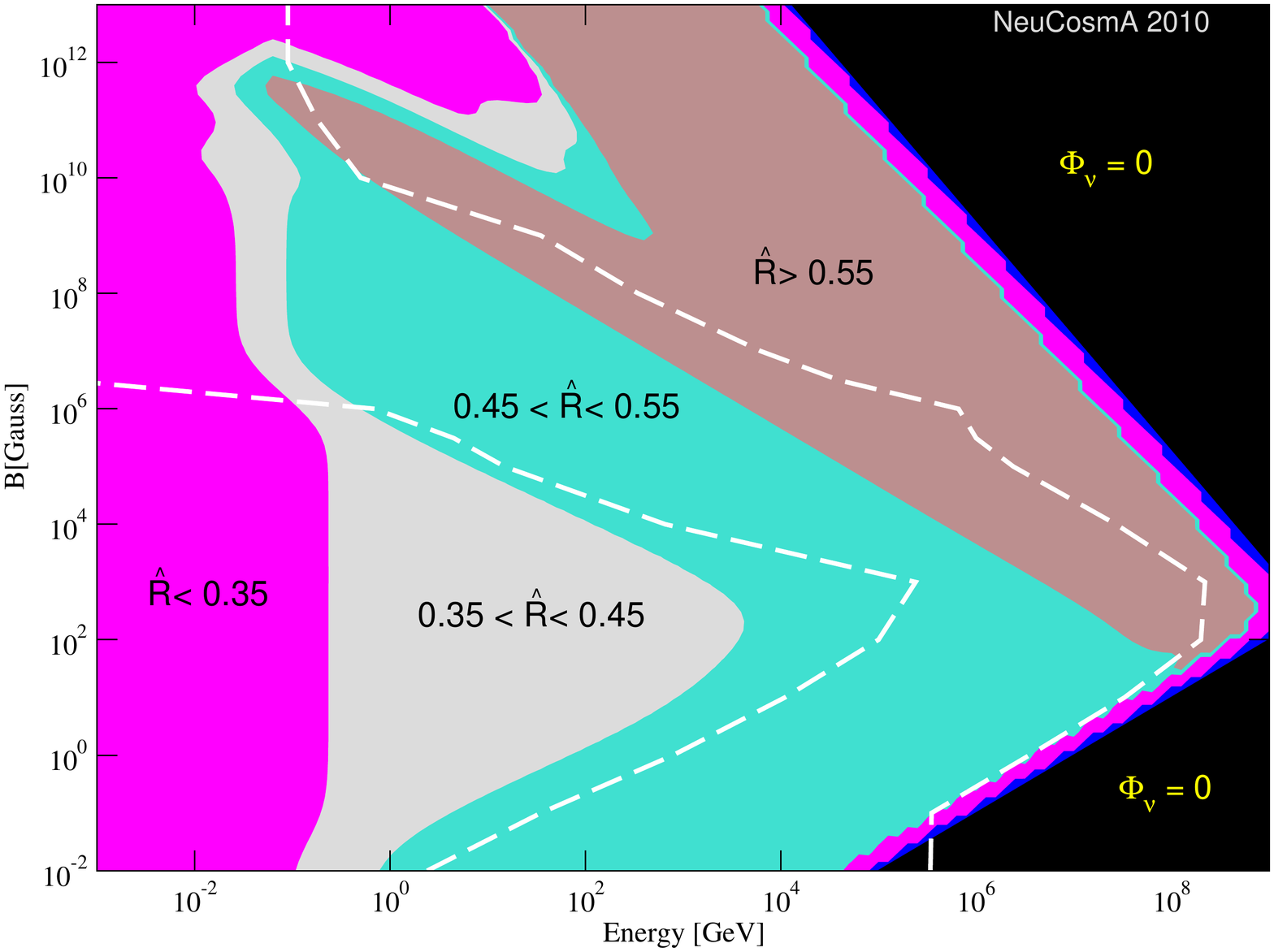}
\mycaption{\label{fig:scan} Contour regions of $\hat R$ in the plane of energy versus magnetic field for $\alpha=2$ and $R=10^{9.7}$ km. The best fit values of the neutrino mixing angles were used. The region between the two dashed white curves is the energy classification range (see text). In the black regions (right corners) the neutrino fluxes go to zero so $\hat R$ is not well-defined.}
\end{figure}
\figu{scan} illustrates the variation of $\hat R$ with the energy and  the magnetic field for $\alpha=2$, $R=10^{9.7}$ km (corresponding to TP3), and the best fit values of the neutrino mixing angles. Four regions are explicitly shown: $\hat R<0.35$, $0.35<\hat R<0.45$, $0.45<\hat R<0.55$ and $\hat R>0.55$. In addition, the energy classification range is shown by the dashed curves. From the figure we see that for energies around $1~\mathrm{TeV}$, $\hat R$ is typically larger than $0.45$ independently of $B$. It is also observed  that the region $\hat R>0.55$ is characterized by large magnetic fields in the energy range of interest for neutrino telescopes. Thus, if $\hat R$ were measured to be smaller than $0.55$ for energies between $1$ and $10$ TeV, we could deduce, from  this figure, that $B\lesssim 10^6~G$. Clearly, it is possible to obtain some information about astrophysical parameters from flavor ratio measurements.   

\begin{figure}[tp]
\begin{tabular}{cc}
\includegraphics[width=0.49\textwidth]{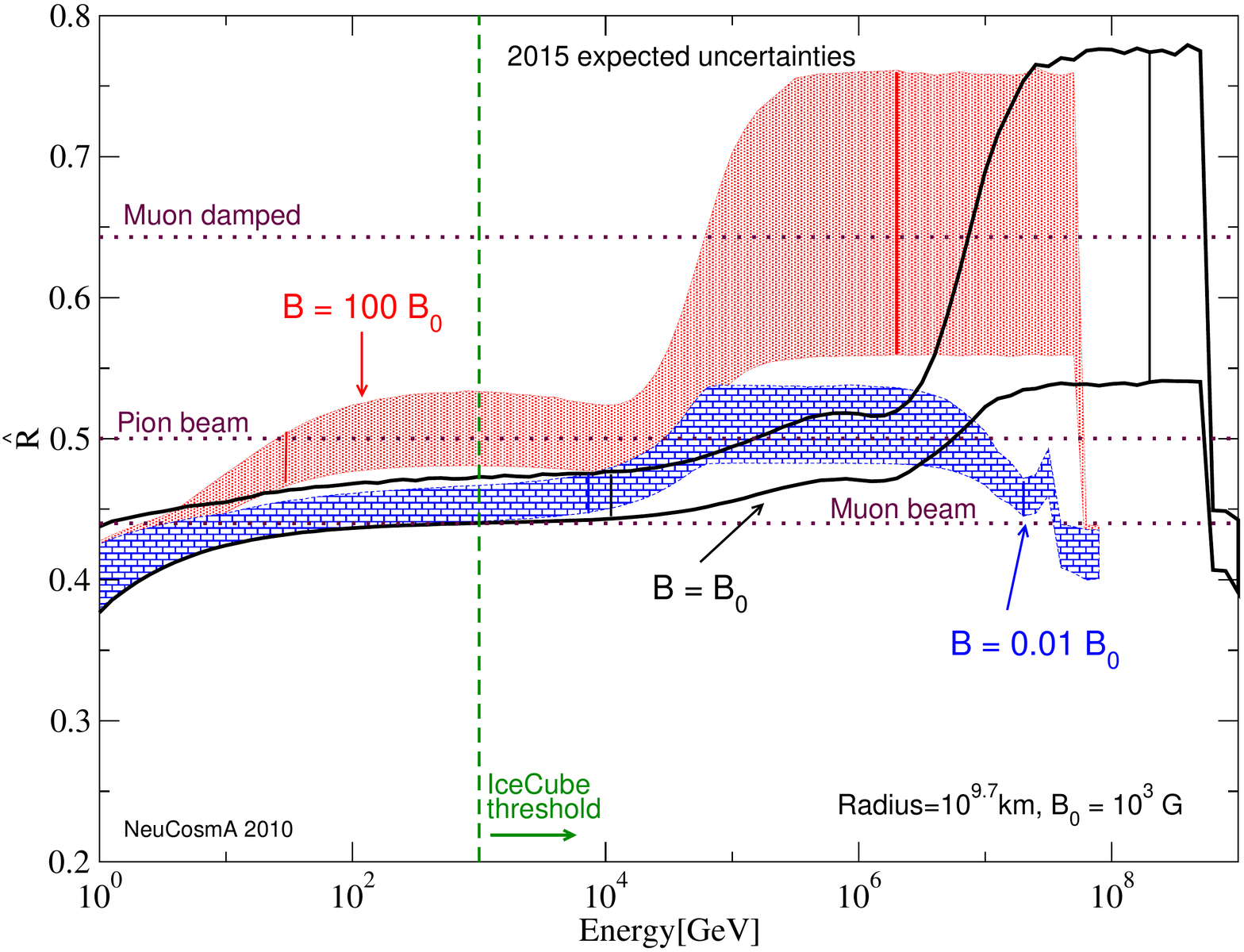} & \includegraphics[width=0.49\textwidth]{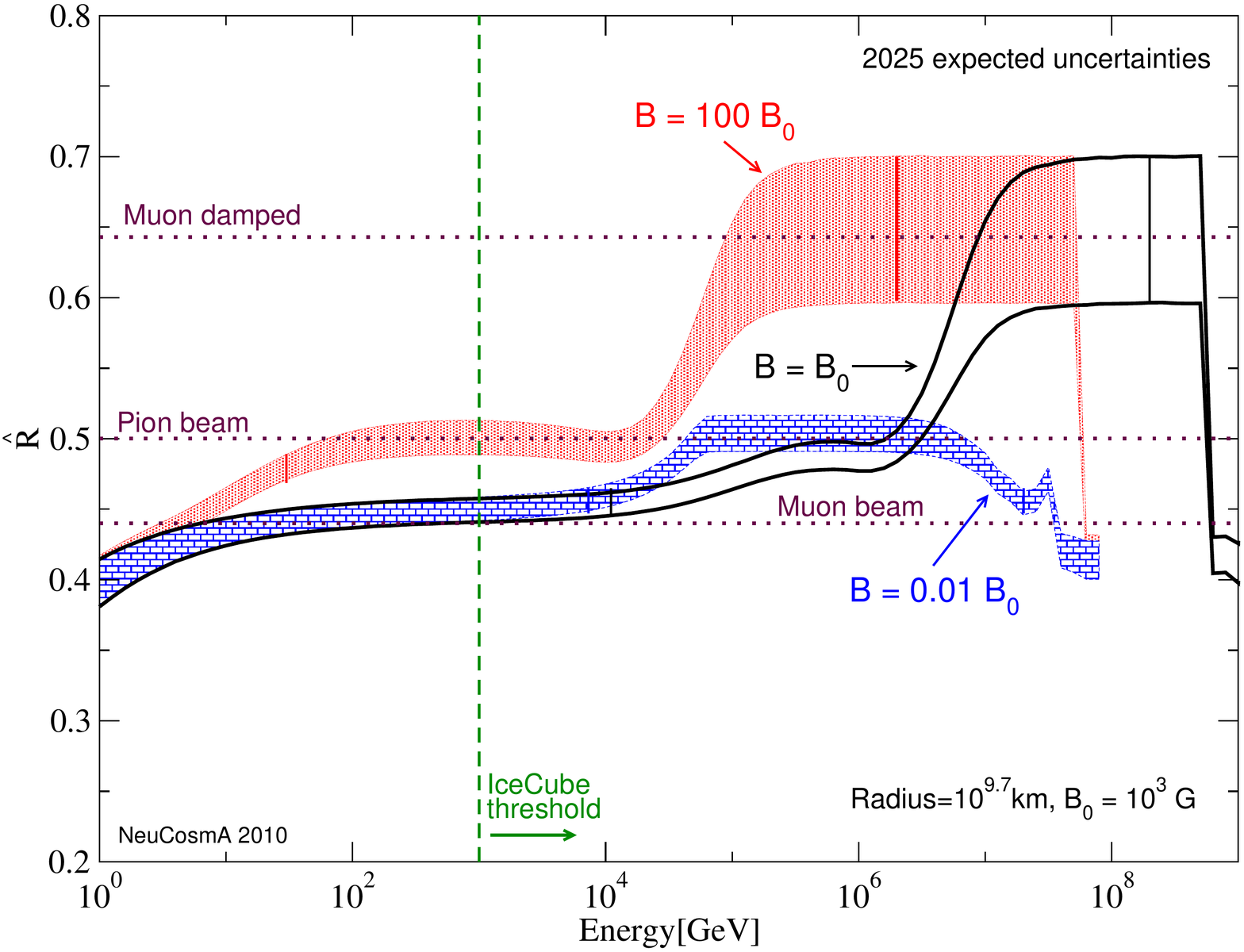}
\end{tabular}
\mycaption{\label{fig:r3Bsvf} $\hat R$ as a function of the energy for three different values of the magnetic field. The bands indicate the expected uncertainty in 2015 (left) and in 2025 (right). The vertical lines mark the energy classification windows (see main text).}
\end{figure}

It is important to analyze if this conclusion still holds when the uncertainties in the neutrino mixing angles are taken into account. To that end, we show in \figu{r3Bsvf} $\hat R$ as a function of the energy for three different values of $B$. As before we take $\alpha=2$ and $R=10^{9.7}$ km, but we now include the  expected uncertainty in neutrino mixing parameters for 2015 (left) and 2025 (right). From the figures, we notice that by measuring $\hat R$ over an energy range above $1~$TeV, one could indeed gain some information about the magnetic field. If $B=10^5$ G (upper shaded/red region), for instance, one expects a transition from a pion beam source to a muon damped source for energies between $10$ and $100$ TeV. But if $B=10$ G (lower shaded/blue region), no transition is expected and $\hat R$ is never larger than about $0.5$. We see then that the expected uncertainties in the mixing angles do not spoil the possibility of using $\hat R$ to obtain information on the magnetic field of cosmic accelerators. However, the information from 2015 is on the borderline on being useful, since the regions, which come from very different magnetic fields, somewhat overlap.

\subsection{Optically thin $\boldsymbol{p\gamma}$ sources?}

In order to test the Glashow resonance at about 6.3~PeV, which is sensitive to electron antineutrinos only, it is useful to consider the flavor ratio
\begin{equation}
\hat T \equiv \frac{\phi_{\bar \nu_e}^{\mathrm{Det}}}{\phi_{\nu_\mu}^{\mathrm{Det}}+\phi_{\bar\nu_\mu}^{\mathrm{Det}}}
\end{equation}
at the detector, representing the number of Glashow resonance versus muon track events. The fluxes at the detector, of course, include flavor mixing, which depends on the mixing parameters. Therefore, we need to include the uncertainties on the mixing parameters. 

\begin{table}[t!]
\begin{center}
\begin{tabular}{lccc}
\hline
 & Pion beam & Muon damped & Muon beam \\
\hline
\multicolumn{4}{l}{\bf{$\boldsymbol{\hat{T}}$ for best-fit values }} \\
Ideal $pp$ source & 0.50 & 0.28 & 0.64 \\
Ideal $p\gamma$ source & 0.22 & 0.00 & 0.36 \\
p$\gamma$ source with 20\% $\pi^-$ contamination & 0.31 & 0.09 & 0.51 \\
\hline
\multicolumn{4}{l}{\bf{$\boldsymbol{\hat T^{\mathrm{min}}}$ for ideal $\boldsymbol{pp}$ source (marginalized over oscillation parameter ranges)}} \\
Current bounds &  0.38 & 0.15 & 0.56 \\
2025 & 0.47 & 0.23 & 0.62  \\
\hline
\end{tabular}
\end{center}
\mycaption{\label{tab:tv} Values of flavor ratio $\hat{T}$ used for ``$p\gamma$ optically thin'' analysis (see main text for details).  }
\end{table}

The purpose of this process can be the discrimination between $p\gamma$ optically thin sources, which produce more $\pi^+$ than $\pi^-$, and $pp$ or optically thick sources, which produce comparable amounts of $\pi^+$ and $\pi^-$. Even after flavor mixing, the value of $\hat T$ is much smaller for a $p \gamma$ optically thin source than any of the other mentioned sources. For instance, for the ideal muon damped $p\gamma$ source, $\hat T \simeq 0$ because no antineutrinos (of either flavor) are produced at the source. In order to reject the hypothesis of a source symmetric in $\pi^+$-$\pi^-$ production, we compute the theoretical minimal possible value $\hat T^{\mathrm{min}}$ for the $\pi^+$-$\pi^-$ {\em symmetric} source after flavor mixing for a pion beam, muon damped source, and muon beam source separately, assuming that $\pi^+$ and $\pi^-$ are produced in equal amounts at the source. For that we vary the oscillation parameters in their $3 \sigma$ allowed ranges, or use the bounds from 2025 (these values correspond to the lower end of the marginalized $\hat T$).  The resulting values can be read off from \Tab~\ref{tab:tv}. In the table, we also show the $\hat{T}$ for the used best-fit values for an ideal $pp$ source ($\pi^+$-$\pi^-$ symmetric) and ideal $p\gamma$ source ($\pi^+$ only) expected from the $\Delta$-resonance approximation only (\cf, \equ{ds}), as well as for the a $p\gamma$ source with a 20\% $\pi^-$ contamination, as it can be expected from high energy processes contributing to the photohadronic interactions.

In order to identify an optically thin $p\gamma$ source, we require $\hat T<\hat T^{\mathrm{min}}$ at the detector for $E=6.3 \, \mathrm{PeV}$. Note that we do not include statistics and systematics in this discussion, we only discuss the theoretical possibility to use the Glashow resonance. From \Tab~\ref{tab:tv}, we can read off that even for the current bounds within each source type, $\hat T<\hat T^{\mathrm{min}}$ even for the $p\gamma$ sources with a 20\% contamination of $\pi^-$ (compare the values from the third and fourth rows filled with numbers). Therefore, a source identification is, in principle, theoretically possible. This possibility is so far based on the assumption that the source can be identified and that the flavor ratio at the source is ``ideal'' for each specific source.  It is obvious from the values for $\hat T^{\mathrm{min}}$ that the identification of the source is mandatory at least for pion beam and muon beam sources. For example, assume that $\hat T \simeq 0.3$ is measured without source identification. In this case, it can be interpreted as muon damped $pp$, or pion beam $p \gamma$ source. Therefore, we require that the type of source (pion beam, muon damped or muon beam) can be identified from the flavor ratio with the same energy ranges (at the source) as discussed in \Sec~\ref{sec:source}, over at least one order of magnitude. However, we choose the muon neutrino flux at the detector to determine the relevant energy classification range. 

For each source type within its allowed energy range in $E_{\mathrm{source}}$, we compute the Lorentz boost factor range from $\Gamma \simeq 6.3 \, \mathrm{PeV}/E_{\mathrm{source}}$ for which the Glashow resonance is matched and for which $\hat T<\hat T^{\mathrm{min}}$, \ie, the optically thin $p \gamma$ source can be identified. Note that compared to \Sec~\ref{sec:pgsource}, we focus on one particular energy (at the detector) for the $p \gamma$ classification. This particular energy, however, corresponds to a range $\Gamma^{\mathrm{\min}} \lesssim \Gamma \lesssim \Gamma^{\mathrm{\max}}$ for which a particular source can be used. Finally, we require $1 \le \Gamma \le 1000$ for reasonable sources. Note that neutron beams are not useful for this type of analysis, since they are not produced by $p\gamma$ interactions. However, whatever effect destroys the $\pi^+$-$\pi^-$ asymmetry, makes $\hat T$ increase, \ie, acts in the same direction.

\begin{figure}[tp]
\centering
\includegraphics[width=\textwidth]{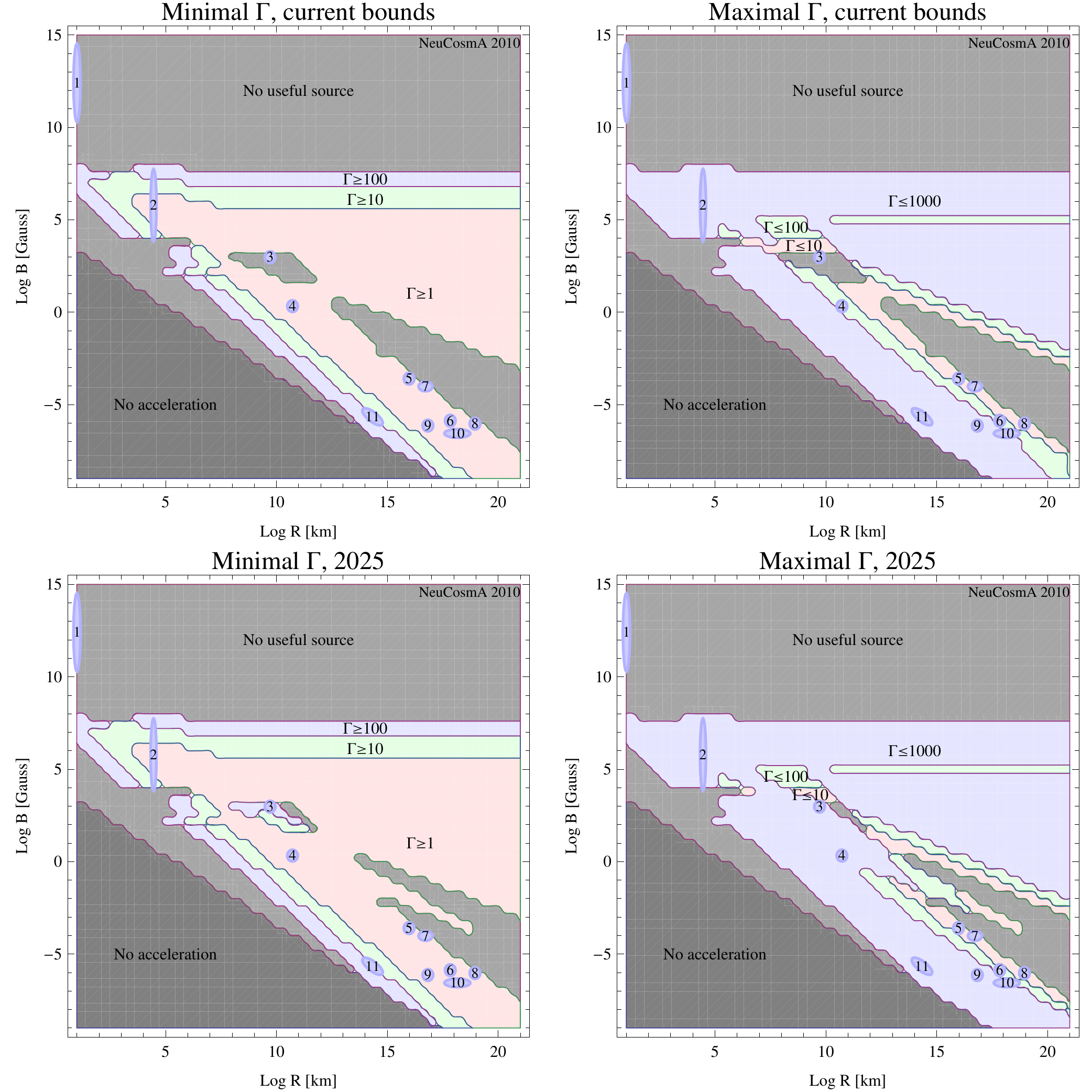}
\mycaption{\label{fig:hillast} Classification of optically thin $p \gamma$ sources from the flavor ratio $\hat T$  in the Hillas plot for $\alpha=2$. The upper row is for the current bounds of the neutrino mixing parameters, the lower row for the 2025 case. The left column shows the minimal $\Gamma$ of the source allowed for matching the Glashow resonance energy, the right column the maximal $\Gamma$ (see main text for details).}
\end{figure}

We show the result of this analysis for $\alpha=2$ in \figu{hillast}.  The left column shows the minimal allowable $\Gamma$ and the right plot the maximal allowable $\Gamma$ within the analysis range. The upper row is for the current bounds, the lower row for 2025. If there are several possible regions which can be used for the Glashow resonance (such as in the pion beam $\rightarrow$ muon damped case), we choose the smallest and highest $\Gamma$ from these regions. This figure (upper panels) is to be interpreted as follows: For example, for test point~6, the Glashow resonance can be used for $1 \lesssim \Gamma \lesssim 100$, which is a reasonable range since these sources will be found around $\Gamma \sim 1$. The same argument applies to test points~4 to~10 for different maximal allowed values of $\Gamma$, which are the typical pion beam sources which can be used for this type of analysis, and test point~2 as muon beam $\rightarrow$ muon damped source. Note that for test point~4, for which the Lorentz boost is often larger than~10, values as large as~1000 are also fine. Since the neutrino energies for test points~11 are lower, higher boost factors $\Gamma \gtrsim 10$ would be required, which might be unrealistic for this type of source.  

In \figu{hillast}, there are several gray-shaded regions marked ``no useful source'' where the Glashow resonance cannot be used. The reason is different in different parts of the parameter space. For large values of $B$, we have neutron beam sources or mixed sources, for which the discrimination between $pp$ and $p\gamma$ interactions is not applicable. Around $B \sim 10^8 \, \mathrm{G}$, the synchrotron cooling prevents neutrino energies high enough, which means that $\Gamma \gg 1000$ would be required. The same applies to the region around $R \sim 10^3 \, \mathrm{km}$ and $B \sim 10^3 \, \mathrm{G}$. Around test point~3 and at the spike at $R \sim 10^{17} \, \mathrm{km}$ and $B \sim 10^{-3} \, \mathrm{G}$ the energies in the identified ranges are typically too high to match the Glashow resonance. Only for improved bounds on the mixing parameters (lower panels), a muon beam range at lower energies can be found in some part of the parameter space, where the source can be identified and the Glashow resonance is matched together with $\hat T<\hat T^{\mathrm{min}}$ in a certain $\Gamma$ range. For test point~3, the required $\Gamma$ factors are still very high.
Note that the energy classification window is slightly different than in \Sec~\ref{sec:pgsource} here.


\section{Summary and conclusions}
\label{sec:summary}

We have considered neutrino production by photohadronic interactions in cosmic accelerators. Our model assumes that protons interact with photons from co-accelerated electrons and positrons. We have included magnetic field effects, such as synchrotron cooling of the charged particle species, and flavor effects, such as the helicity dependent muon decays affecting the flavor ratios at the source and flavor mixing. In addition, we have considered neutrinos from kaons and neutrons produced in the photohadronic interactions. Because the spectral shapes, flavor ratios,  and neutrino-antineutrino ratios, which we have investigated, strongly depend on the initial charged pion distributions coming from the photohadronic interactions, we have used an efficient method to compute the secondary particle spectra based on \Ref~\cite{Hummer:2010vx} parameterizing the physics of SOPHIA. This approach includes direct production and high energy processes and allows for a computation of one model within a few seconds. We have used this approach for a large scale parameter space scan as a function of $R$ (size of the acceleration region) and $B$ (magnetic field) to study different sources as a function of the Hillas plot parameters. Another important parameter of our model is the (universal) injection index $\alpha$.

One of the quantities of interest has been the ratio between electron and muon neutrinos (``flavor ratio'') at the source, which characterizes its flavor composition before propagation to the Earth. For example, ``pion beam sources'' produce neutrinos in the flavor ratio $\nu_e$:$\nu_\mu$:$\nu_\tau$ of 1:2:0, ``muon damped sources'' in the ratio 0:1:0 (the muons loose energy in the magnetic field before they decay), and ``neutron beam sources'' in the ratio 1:0:0 (the neutrinos are produced by beta decays).  As a new type of source, we have found  ``muon beam sources'' with the ratio 1:1:0, similar to a neutrino factory. In this case, the muons from higher energies pile up in the lower energy part of the spectrum, where pion decays hardly contribute.
We have illustrated that a discussion of the flavor ratio only makes sense if one considers the corresponding spectrum simultaneously, since the region around the spectral peak (in $E^2 Q_\nu$) will contribute most to the event rates. Therefore, we have focused on an energy range around the peak for the source classification, and we have required a unique classification over at least one order of magnitude in energy. For $\alpha=2$, our main result can be found in \figu{hillas}. 
 Interpreting the test points in this figure in terms of the sources listed there, most sources on galactic scales perform as the classical pion beam. Neutron stars, on the other hand, are typical neutron beams because the strong magnetic field damp all charged particle species. However, the neutrino energies are probably too low to be detected. The most interesting regions may be around white dwarfs (TP~2) and active galactic nuclei (TP~3). For example, active galactic nuclei (TP~3) are classified as muon damped sources for $\alpha=2$ because neutron decays and magnetic field effects prohibit the classification of the pion beam energy range. For $\alpha=3$, however, they show both pion beam and muon damped ranges, and for even softer sources they perform as pion beams.  The white dwarfs (TP~2), however, show for $\alpha=2$ a muon beam $\rightarrow$ muon damped behavior, \ie, the muon neutrino spectra exhibits a spectral split coming from the muon damping.
Although the target photon field of GRBs cannot be described within our model, we have shown that for pion spectra characteristic in GRBs the characteristic pion beam $\rightarrow$ muon damped energy dependence from \Ref~\cite{Kashti:2005qa} is reproduced. Apart from the discussed flavor ratios, fully anomalous flavor ratios are found in some regions of the parameter space especially in regions where strong magnetic field effects compete with neutron decays (the neutrons are not affected by the magnetic fields).

While the above classification has been performed at the source, the situation at the detector is different. First of all, flavor mixing enters as a propagation effect. And second, the acceleration region may be Lorentz boosted with respect to the observer blue-shifting the neutrino energies, whereas the cosmological expansion redshifts the energies. The absolute energies at the detector are, however, relevant, since they have to be above the detection threshold. We have demonstrated that these effects do not, in principle, prohibit the detection of flavor ratios including the source classification for many characteristic test cases. However, the knowledge on the neutrino mixing parameters can only be expected to be good enough after the  T2K and Daya Bay results for serious applications. On the other hand, we have shown for a luminous point source, that some information on the source may be extracted from the flavor ratios. For example, if all the other parameters can be estimated from related measurements, such as the size of the acceleration region by the variability timescale in gamma ray observations, the flavor ratio allows in principle for independent constraints on the magnetic fields of the source.

Finally, we have discussed if the Glashow resonance at 6.3~PeV, which is sensitive to electron antineutrinos only and therefore to the neutrino-antineutrino ratio, can be used to learn something about the neutrino production mechanism. For example, in $p\gamma$ interactions significantly more $\pi^+$ than $\pi^-$ are expected to be produced in sources optically thin to neutrons. This asymmetry can be destroyed by subsequent neutron interactions in optically thick sources, a significant contribution of the high energy photohadronic processes, a significant contribution of $pp$ interactions, or neutron decays leading to a $\bar \nu_e$ typically not present in such sources. We have performed an analysis at both the source and at the detector. While at the source, our ``$p\gamma$ optically thin'' condition is satisfied for most accelerators at least in some energy range, at the detector the Glashow energy has to be exactly matched, and the type of sources has to be identified to disentangle the effects from flavor mixing. This leads to constraints to the possible boost factors $\Gamma$ of the sources. For example, the active galactic nuclei (TP~3) typically cannot be identified as $p\gamma$ optically thin. Sources on galactic scales or larger typically can be easily classified.

We conclude that flavor and neutrino-antineutrino ratios of astrophysical sources may be interesting to study both particle physics and astrophysics properties of the sources. Flavor ratios are an important possibility to test the magnetic fields in the source, because these lead to a spectral split of the neutrino spectra from pion, muon, and kaon decays. On the other hand  the neutrino-antineutrino ratio can be used to study the neutrino production mechanism. Propagation effects between source and detector do, in principle, not void these effects. On the other hand, they may be used to study non-standard neutrino properties, such as decoherence or decays. Therefore, the study of flavor ratios and the Glashow resonance at neutrino telescopes should be higher valued, and future instruments should be optimized for that once cosmic neutrinos are detected.

\subsubsection*{Acknowledgments}

We would like to thank P. Baerwald, K. Mannheim, D. Meloni, I. Mocioiu, S. Pakvasa, and F. Spanier for useful discussions.
SH and WW  would like to thank UAM Madrid for their local hospitality during our research visits in February 2009 and 2010, and MM and CY would like to thank the Institut f{\"u}r Theoretische Physik und Astrophysik at  Universit{\"a}t W{\"u}rzburg for their hospitality during our research visits in June 2009 and June 2010. 
SH acknowledges support from the Studienstiftung des deutschen Volkes (German National Academic Foundation) and from the Research Training Group GRK1147 ``Theoretical astrophysics and particle physics'' of Deutsche Forschungsgemeinschaft. MM is supported by Spanish MICINN grants FPA-2009-08958, FPA-2009-09017 and consolider-ingenio 2010 grant CSD-2008-0037, by CSIC grant 200950I111, and by Comunidad Autonoma de Madrid through the HEPHACOS project S2009/ESP-1473. WW would like to acknowledge support from the Emmy Noether program of Deutsche Forschungsgemeinschaft, contract WI 2639/2-1. CY is supported by the \emph{Juan de la Cierva} program of the MICINN of Spain. He acknowledges additional support from the MICINN Consolider-Ingenio 2010 Programme under grant MULTIDARK CSD2009-00064, from the MCIINN under Proyecto Nacional FPA2009-08958, and from the CAM under grant HEPHACOS S2009/ESP-1473.


\end{document}